\documentclass[journal]{IEEEtran}
\usepackage{algorithm}
\usepackage{multirow}
\usepackage{array}
\usepackage{algpseudocode}
\usepackage{amsmath}
\usepackage{cite}
\usepackage{booktabs}
\usepackage{graphicx}
\usepackage{float}
\usepackage{threeparttable}
\usepackage{color}
\usepackage{tikz}
\usepackage{epstopdf}
\usepackage{amssymb}
\usepackage{graphicx}
\usepackage{subfigure}
\usepackage{url}
\usepackage{setspace}
\newlength{\tempdima}
\newcommand{\rowname}[1]
{\rotatebox{90}{\makebox[\tempdima][c]{\textbf{#1}}}}

\newcommand{\luoma}{\uppercase\expandafter{\romannumeral1}}
\newcommand{\luomas}{\uppercase\expandafter{\romannumeral1}~}
\begin{document}	

\title{Random Directional Attack for Fooling Deep Neural Networks}

\author{Wenjian~Luo,
	Chenwang~Wu,
	Nan~Zhou and
	Li~Ni
	\thanks{This study is supported by the National Natural Science Foundation of China and the open project of the Anhui Province Key Laboratory of Intelligent Building and Building Energy Saving. (\textit{Corresponding author: Wenjian Luo.})}
	\thanks{Wenjian Luo, Chenwang Wu, Nan Zhou and Li Ni are with Anhui Province Key Laboratory of Software Engineering in Computing and Communication, the School of Computer Science and Technology, the University of Science and Technology of China, Hefei 230027, Anhui, China.}	
	\thanks{Email: wjluo@ustc.edu.cn, \{wcw1996, hinanmu, nlcs\}@mail.ustc.edu.cn.}
}

\maketitle
\begin{abstract}
Deep neural networks (DNNs) have been widely used in many fields such as images processing, speech recognition; however, they are vulnerable to adversarial examples, and this is a security issue worthy of attention. Because the training process of DNNs converge the loss by updating the weights along the gradient descent direction, many gradient-based methods attempt to destroy the DNN model by adding perturbations in the gradient direction. Unfortunately, as the model is nonlinear in most cases, the addition of perturbations in the gradient direction does not necessarily increase loss. Thus, we propose a random directed attack (RDA) for generating adversarial examples in this paper. Rather than limiting the gradient direction to generate an attack, RDA searches the attack direction based on hill climbing and uses multiple strategies to avoid local optima that cause attack failure. Compared with state-of-the-art gradient-based methods, the attack performance of RDA is very competitive. Moreover, RDA can attack without any internal knowledge of the model, and its performance under black-box attack is similar to that of the white-box attack in most cases, which is difficult to achieve using existing gradient-based attack methods.
\end{abstract}

\IEEEpeerreviewmaketitle

\section{Introduction}

Nowadays, deep neural networks (DNNs) occupy an unassailable position in many fields of image processing\cite{xie2012image,pathak2016context}, natural language processing\cite{collobert2008unified}, and speech recognition\cite{hinton2012deep}. However, recent work has shown that DNN models, regardless of classification accuracy, are susceptible to misclassification because of perturbation noises that are difficult to detect by the human eye. Further, studies found that the added disturbances might be transferable, i.e., they can simultaneously deceive models with different structures. The mechanism of intentionally adding subtle perturbations to cause a target model to misclassify examples with high confidence is called an adversarial attack; the disturbed example is called an adversarial example\cite{szegedy2013intriguing}.

The discovery of adversarial perturbations attracted a number of researchers to conduct an in-depth study on adversarial learning. Moosavi-Dezfooli et al. \cite{moosavi2017universal} proposed the existence of universal perturbation, i.e., a disturbance that can successfully attack a sample can also be applied to a new sample and to spoof the same network with high probability. Athalye et al. \cite{athalye2017synthesizing} found that real objects generated by adversarial samples through 3D printing techniques could also attack DNN classifiers. Xie et al. \cite{xie2017adversarial} extended the adversarial examples to the field of semantic segmentation; these examples were wrongly segmented or detected. 
In addition to the adversarial attacks in image classification, Carlini et al. \cite{carlini2018audio} stated that advanced speech-to-text models are susceptible to adversarial perturbations, and any waveform given could be transcribed into a specified text by adding an almost inaudible waveform. 
Alzantot et al. \cite{alzantot2018did} misclassified speech by changing the unimportant parts of the speech segments. 
Ebrahimi et al. \cite{ebrahimi2017hotflip} successfully attacked word-level classifiers by atomic flipping operations and retaining semantic constraints.

\begin{figure}
	\centering
	\includegraphics[width=85mm]{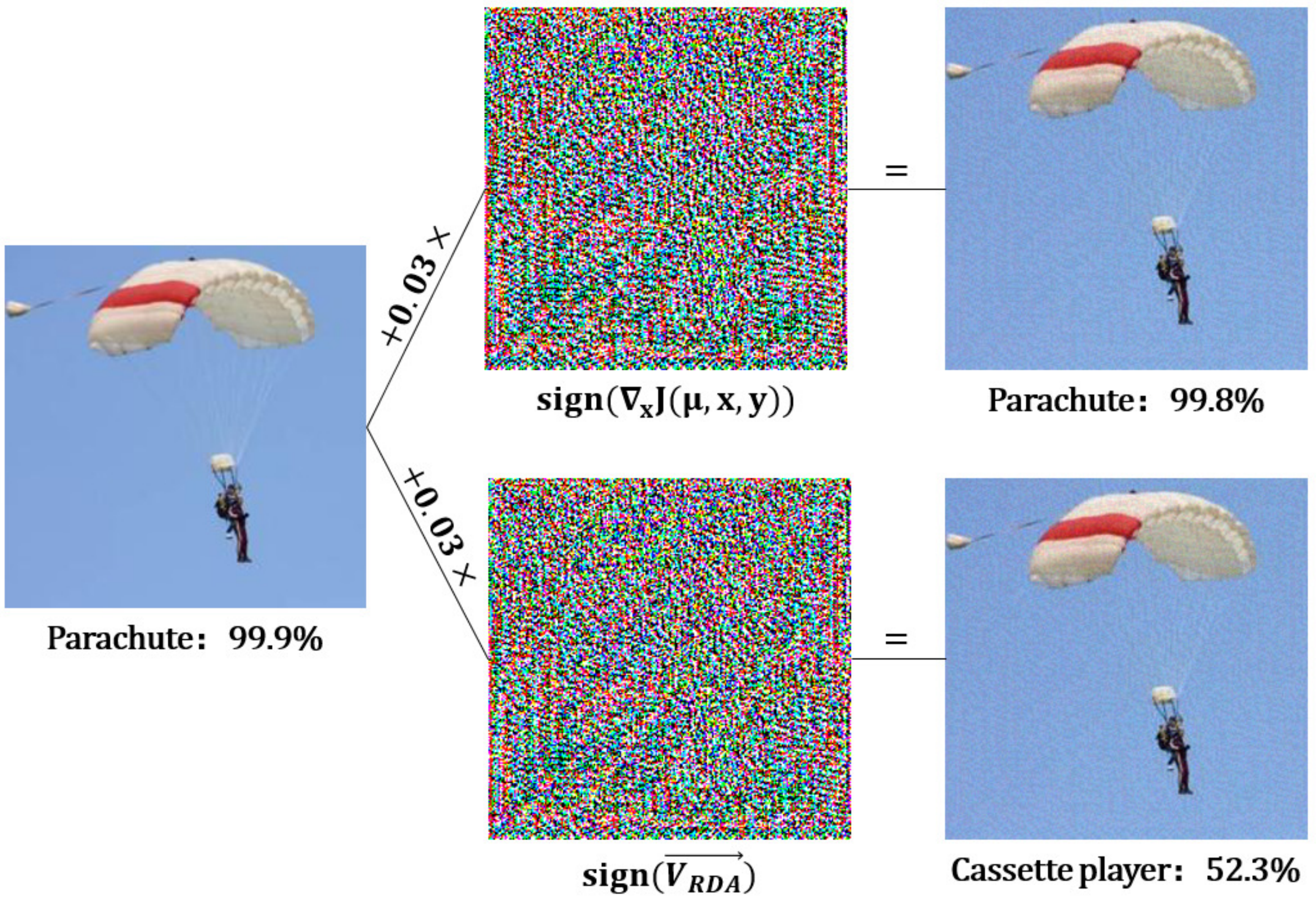}\\
	\caption{Examples of attacks on the same image by FGSM and RDA based on one-step attack. \textbf{Left column}: clean samples to be attacked. \textbf{Middle column}: FGSM attack direction (top) and RDA attack direction (bottom), and their included angle is $36.1^\circ$. \textbf{Right column}: sample generated by FGSM (top) and sample generated by RDA (bottom). FGSM fails, but RDA attacks successfully. Note that in order to reduce the number of iterations, RDA stops when the attack succeeds. Therefore, although the confidence of the error class Cassette player is only 52.3\%, the attack has been successful.}
	\label{compare}
\end{figure}

Existing attack methods can be classified into white-box attacks and black-box attacks based on the amount of knowledge that can be obtained from the network \cite{akhtar2018threat}. In a white-box attack, an attacker can obtain all the information about the network (including model structure, weights, and even training details). In a black-box attack, the attacker cannot obtain all the information of the model; however, they only know the type of data processed by the model and the common interactions performed by the model, such as the outputs of the model. The implementation of the black-box attack utilizes the characteristics of the transferability of the adversarial perturbations, and the attack is performed by delivering the adversarial examples generated by the local model to the target model.
Thus, an effective black-box attack on the target model is often based on an efficient white-box attack. 
Unfortunately, excellent white-box attacks are poorly transferable, while highly transferable methods tend to perform poorly \cite{kurakin2016adversarial}. 
Thus, the performance of existing black-box attacks is generally unsatisfactory\cite{dong2018boosting}.

Training neural networks is a process that minimizes the loss function along the gradient descent direction. Therefore, it is natural to use gradient information to attack the trained model. This type of methods must ensure that the gradient direction of the input is towards the direction of increasing loss under a subtle perturbation; however, the nonlinear model does not necessarily satisfy this condition. 

Because perturbations added along the gradient direction do not necessarily successfully attack the sample, any other directions that can ensure a successful attack should be identified. 

This paper proposes a random directional attack (RDA), which uses hill climbing search to find the direction along which a sample can be successfully attacked under a minor disturbance. RDA uses the first-choice hill climbing algorithm to find a successful direction, where only partial dimensions are randomly selected at each iteration of hill climbing. 
Experimental results on multiple datasets demonstrate that the attack performance of RDA is competitive with state-of-the-art gradient-based attack methods. 
In Fig. \ref{compare}, we show the results of fast gradient sign method (FGSM) \cite{goodfellow2014explaining} and RDA attacking the same sample. FGSM has no effect on the classifier, whereas RDA can cause model misclassification.

The primary contributions of the paper are summarized as follows.
\begin{itemize}
\item Based on the fact that an attack is not necessarily in the gradient direction, we propose RDA, which searches the attack direction based on a random but directional search, i.e., the first-choice hill climbing on random neighbors. It implements a one-step attack on the direction that is identified. 
\item RDA demonstrates its excellent attack performance in white-box attacks. In particular, as a one-step attack method, RDA is significantly better than the existing FGSM.
Although RDA uses the gradient of the target model as the starting point under white-box attacks, it could use the gradient of the substitute model under black-box attacks, and it achieves a similar performance to that of white-box attacks.
\item The excellent attack performance of RDA confirms that the effective attack direction may deviate from the gradient direction. We found that, for samples difficult to be perturbed in the gradient direction, the effective attack directions may deviate considerably from the gradient direction.
\item The experimental results show that the angles included between the effective attack direction and the gradient direction vary considerably, and there are no explicit laws. This indicates that the adversarial region is randomly and irregularly distributed. Thus, it is of great significance and a great challenge to form a unified theory to explain the adversary in DNNs.
\end{itemize}

The rest of this paper is organized as follows. Section \ref{sec:relatedwork} introduces the vulnerability of DNNs and the specific work of attacking and defending. Section \ref{sec:methodology} elaborates on the ideas of the random directional attack and the procedure of the algorithm. Section \ref{sec:experiment} verifies the feasibility of RDA through experiments. Finally, the last section provides a summary of the paper.

\section{Related Work}
\label{sec:relatedwork}
In this section, we first discuss the vulnerability of DNNs. Second, we list some classical methods for generating adversarial examples. Finally, we introduce some studies related to defending against these adversarial examples.

\subsection{Vulnerability of DNNs}
Considerable work has been invested in studying the vulnerability of DNNs, i.e., why do there exist so many adversarial examples. 
Szegedy et al. studied the adversarial examples in 2013\cite{szegedy2013intriguing}.
At that time, the general view was that the highly nonlinear characteristics of DNNs were the main reason for the existence of adversarial examples. DNNs could not fully learn the completely generalized features of data, and therefore, these perturbed examples were randomly and low-probably distributed in a high-dimensional network model space\cite{szegedy2013intriguing}. However, Goodfellow et al. \cite{goodfellow2014explaining} argued that adversarial perturbations are attributed to the fact that the neural network is too linear, such as LSTM and ReLU, which are intentionally designed to be very linear. Tanay et al. \cite{tanay2016boundary} questioned the linear interpretation of the network and proposed that it is easy to generate adversarial examples when the decision boundary of the classifier is located near the submanifold of the data. According to experimental comparisons of various denoising algorithms, Gu et al. \cite{gu2014towards} concluded that the vulnerability of DNNs has very little to do with the structure of the selected model; however, it is related to the training process and the defects of the objective function used. Tabacof et al. \cite{tabacof2016exploring} used various noises to probe the pixel space of the image and concluded that adversarial perturbations are prone to occur in a large area of the pixel space. Cubuk et al. \cite{cubuk2017intriguing} attributed the occurrence of adversarial perturbations to the inherent uncertainty of the statistics of the logit differences of model, and the uncertainty does not change significantly in learning, which makes different neural networks have different robustness. Through the study of adversarially robust generalization, Schmidt et al. \cite{schmidt2018adversarially} showed that the robustness of the model strongly depends on the distribution of the data; however, it is independent of the specific category of training, and the standard linear classifier is more robust than the others. Bubeck et al. \cite{bubeck2018adversarial} argued that the reason for the occurrence of perturbations in a high-dimensional classifier is because of the computational limitation of the learning algorithms. 
Ilyas et al. \cite{ilyas2019adversarial} proposed that adversarial examples originate from the existence of non-robust features that are difficult for humans to understand and capture in experimental datasets. Stutz et al. \cite{stutz2019disentangling} demonstrated that the adversarial perturbations exist both outside the manifold and in the manifold, and they proposed that adding the adversarial examples in the manifold to training data is helpful in improving the generalization ability of the model.

\subsection{Adversarial Attack}
An adversarial attack finds a minor perturbation that can cause sample misclassification. We can transform it into an optimization problem. Let the classifier be $F$, the image to be attacked is $x$, the correct category of the image is $y$, and the added disturbance is $\mathcal{L}$; then, the adversarial attack can be regarded as the optimal solution to solve the following problem.

$$\min \limits_{\mathcal{L}} \left\lVert\mathcal{L}\right\lVert \quad s.t.\ F(x+\mathcal{L})\ne y,x+\mathcal{L}\in \mathbb{R}^m.$$

If $F(x+\mathcal{L})$ only needs to be different from the original category $y$, we call such an attack a non-targeted attack, but if it is a pre-specified category, it is called a targeted attack.

Next, we introduce several gradient-based attack strategies similar to the algorithm in this paper.

\textbf{Fast Gradient Sign Method (FGSM)}\cite{goodfellow2014explaining}: Training neural networks is a process that minimizes the loss function. If we find the gradient of the model to the input, adding a small perturbation $\mathcal{L}$ in the gradient direction to increase the loss value will most likely result in a change in the classification.
$$\mathcal{L}=\epsilon\cdot sign(\bigtriangledown_x J(\mu,x,y)),$$
where $\epsilon$ is the threshold for adding the perturbation, $J(\mu,x,y)$ is the loss of the network with input $x$, and the $sign$ function is used to indicate the specific direction of the gradient.

\textbf{Basic Iterative Method (BIM)}\cite{kurakin2016adversarial2}: This is a multiple iteration version of FGSM, which applies the FGSM algorithm multiple times and crops the pixel values after each iteration to meet the perturbation limits in a reasonable range. Let $x^*_t$ denote the adversarial example generated by the $t$-th iteration, and then,
$$x^*_{t+1}=clip_{x,\epsilon}(x^*_t+\alpha\cdot sign(\bigtriangledown_xJ(\mu,x^*_t,y))).$$

\textbf{Iterative Least-likely Class Method (L.L.Class)}\cite{kurakin2016adversarial2}: Similar to BIM, it uses iterative FGSM; however, the algorithm aims to classify the image into the least likely class $y_{LL}=argmin(F(x))$. The main idea of the algorithm is to maximize $log F_{y_{LL}}(x)$ by adding perturbations in the negative gradient direction $-sign(\bigtriangledown_xJ(\mu,x,y_{LL}))$:

$$x^*_{t+1}=clip_{x,\epsilon}(x^*_t-\alpha\cdot sign(\bigtriangledown_xJ(\mu,x^*_t,y_{LL}))).$$

\textbf{Momentum Iterative Fast Gradient Sign Method (MI-FGSM)}\cite{dong2018boosting}: A one-step perturbation in the gradient direction may not achieve the purpose of an attack, and the simple iterative FGSM may fall into a local optimum; therefore, the momentum is integrated into FGSM to maintain a stable direction change of the perturbation.

$$g_{t+1}=\gamma\cdot g_t+\frac{\bigtriangledown_xJ(\mu,x^*_t,y)}{\left\lVert\bigtriangledown_xJ(\mu,x^*_t,y)\right\lVert_1},$$
$$x^*_{t+1}=x^*_t+\epsilon\cdot sign(g_{t+1}),$$
where $g_t$ collects the gradient of the first $t$ iterations, and if the attenuation factor $\gamma$ is 0, the algorithm degenerates to BIM.

In addition to the above gradient-based attacks, Moosavi-Dezfooli et al. \cite{moosavi2016deepfool} estimated the decision boundary of the classifier through multiple iterations and generated the perturbation by calculating the distance. Papernet et al. \cite{papernot2016limitations} attacked the sample by maintaining a saliency map and modifying only a small number of pixels. Carlini et al. \cite{carlini2017towards} designed three attack methods by limiting the $L_0$, $L_2$, and $L_\infty$ norms. Su et al. \cite{su2019one} only modified one pixel to successfully attack the image through differential evolution. Chen et al. \cite{chen2018ead} attempted to solve the problem of adversarial attacks by describing it as an elastic network regularization optimization problem. Xie et al. \cite{xie2019improving} performed a random transformation of the input image to maximize the loss function of the network for the attack. Hosseini et al. \cite{hosseini2018semantic} proposed the concept of a semantic adversarial example, which may have a large number of modifications to the pixel but can maintain the original semantic features of the image. Inspired by semantic preservation, Peng et al. \cite{peng2018structure} proposed using structure-preserving transformation to generate semantic adversarial samples.

\subsection{Adversarial Defense}
The security of DNNs is an important concern that needs to be resolved \cite{akhtar2018threat}. It is natural to think of incorporating the adversarial samples into the training set to allow the model to learn the characteristics of the perturbations\cite{goodfellow2014explaining}, but the effect is very limited. Inspired by network distillation, Papernot et al. \cite{papernot2016distillation} proposed using a distillation technology to hide the model gradient, and thus defend against gradient-based attacks. An ensemble of the networks was shown to have significant effects on improving the robustness of the model\cite{caruana2004ensemble, dong2018boosting}. Bhagoji et al. \cite{bhagoji2017dimensionality} used data dimension reduction and principal component analysis to enhance the robustness of the model. Buckman et al.\cite{buckman2018thermometer} proposed a thermometer code to resist adversarial attacks by modifying the network structure. Samangouei et al. \cite{samangouei2018defense} proposed the use of generative adversarial networks to eliminate the noise from potentially dangerous samples. Jia et al.\cite{jia2019comdefend} proposed a ComDefend model for end-to-end image compression and reconstruction of images for defense. There are also some defense methods based on denoising that attempt to eliminate the effects of perturbations\cite{moosavi2018divide,chen2018image,prakash2018deflecting}.

Many detection methods are based on the statistical knowledge to find anomalies of adversarial samples \cite{feinman2017detecting,grosse2017statistical,ma2018characterizing}. Lu et al.\cite{lu2017safetynet} designed an RBF-SVM to detect the existence of adversaries by studying the different characteristics of the adversarial images and clean images after passing through the ReLU layer. Liang et al. \cite{liang2018detecting} determined whether the sample is disturbed by comparing the image entropy before and after adaptive denoising for the given sample. Lee et al.\cite{lee2018simple} used Mahalanobis distance to define the scores for class-like Gaussian distributions to distinguish adversarial samples and non-distribution samples.

\section{Methodology} 
\label{sec:methodology}

\subsection{Main Idea}
\begin{figure*}
	\settoheight{\tempdima}{\includegraphics[width=.18\linewidth]{example-image-a}}%
	\centering\begin{tabular}{@{}c@{}c@{}c@{}c@{}c@{}c@{}}
		\includegraphics[width=.18\linewidth]{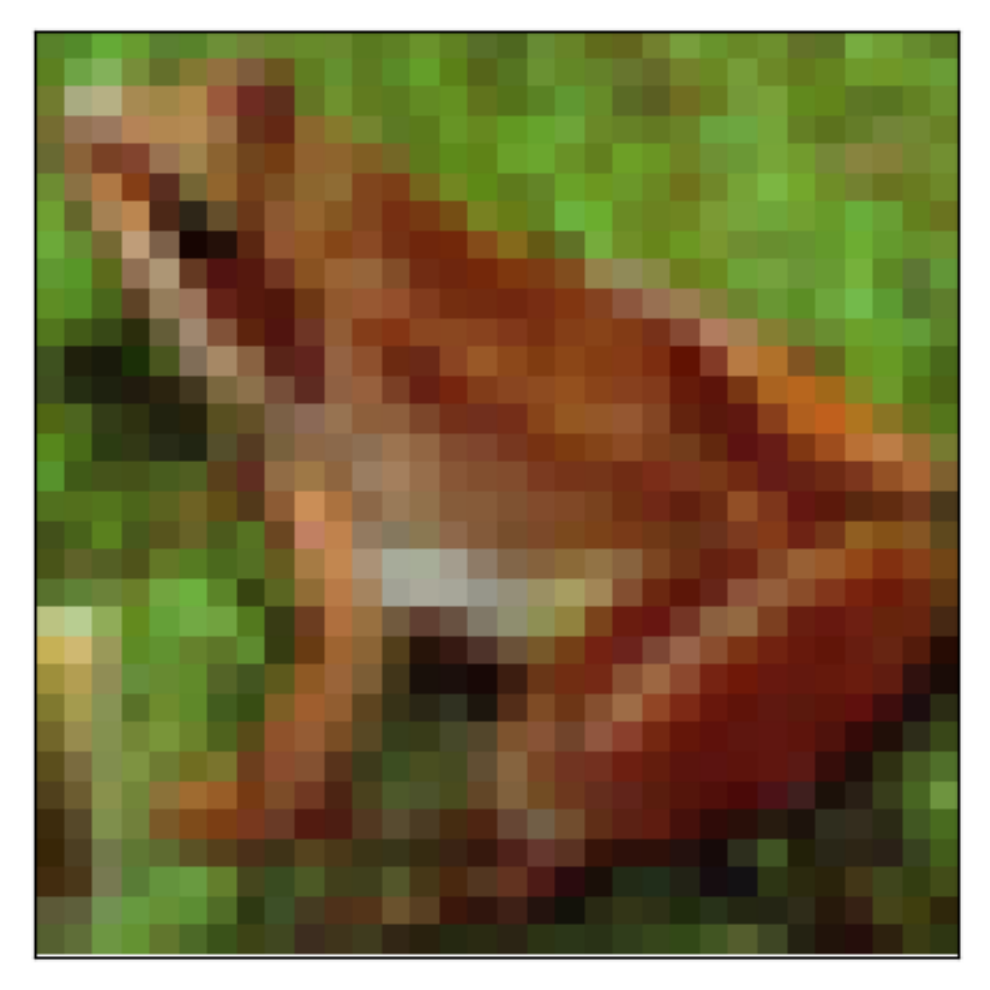}&
		\includegraphics[width=.18\linewidth]{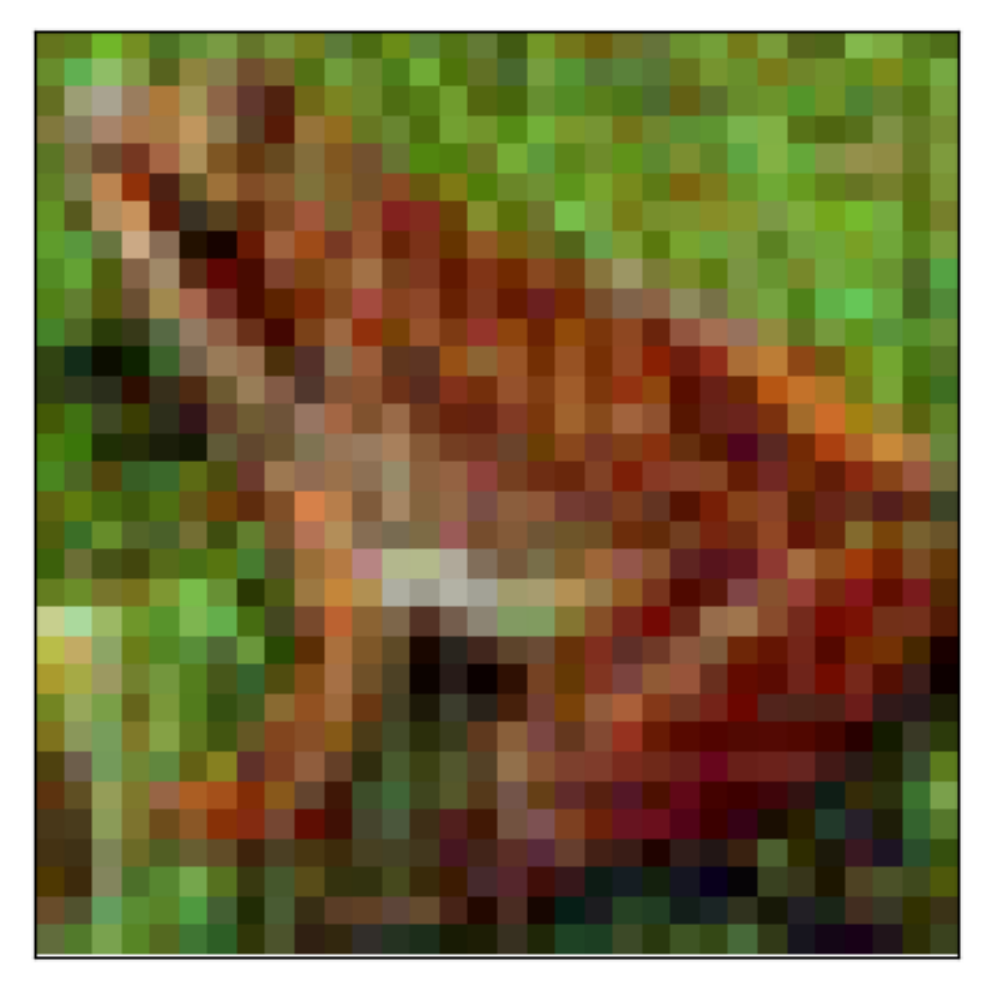}&
		\includegraphics[width=.18\linewidth]{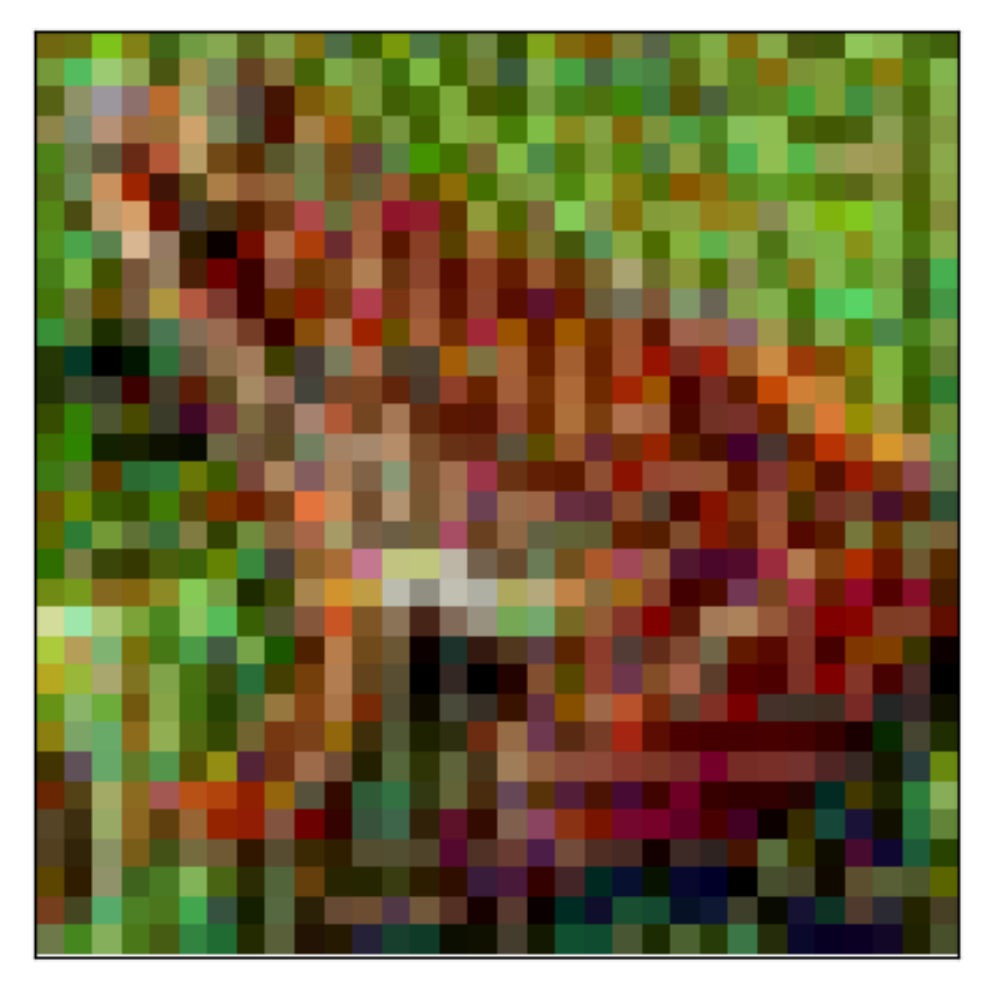}&
		\includegraphics[width=.18\linewidth]{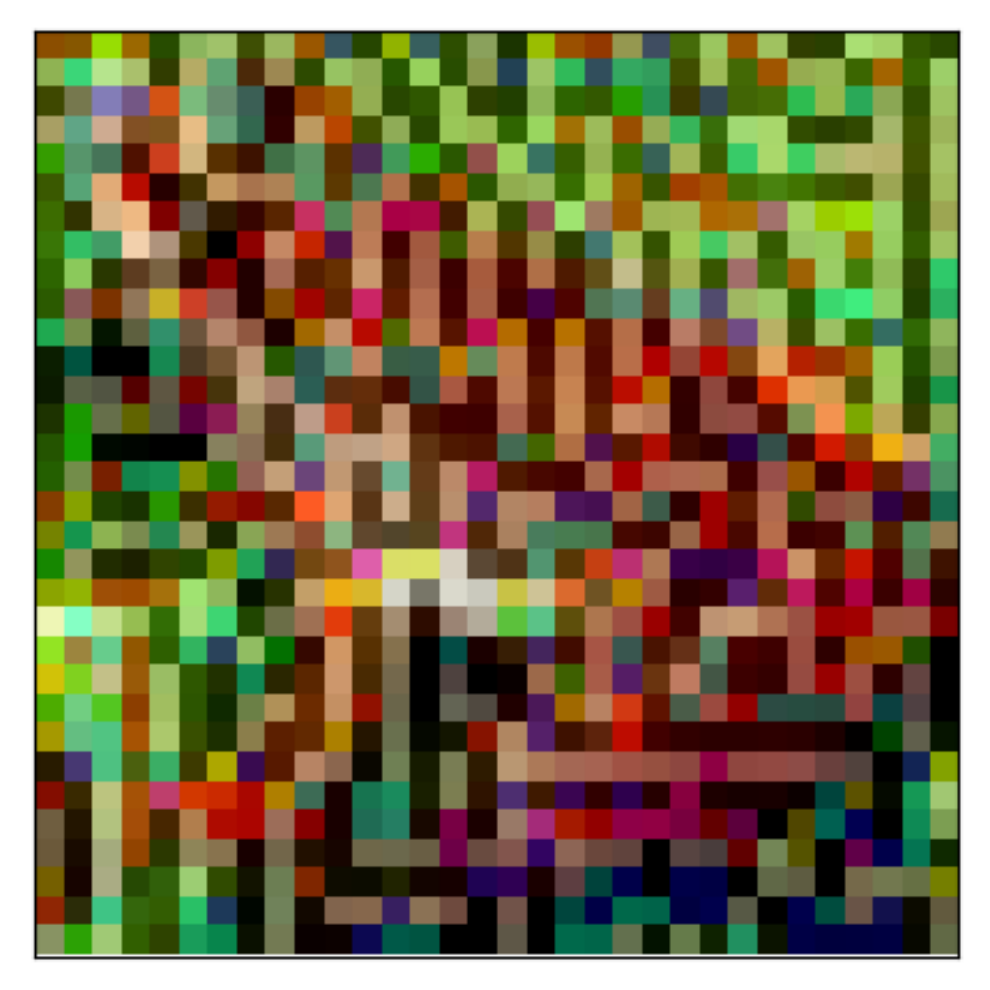}&
		\includegraphics[width=.18\linewidth]{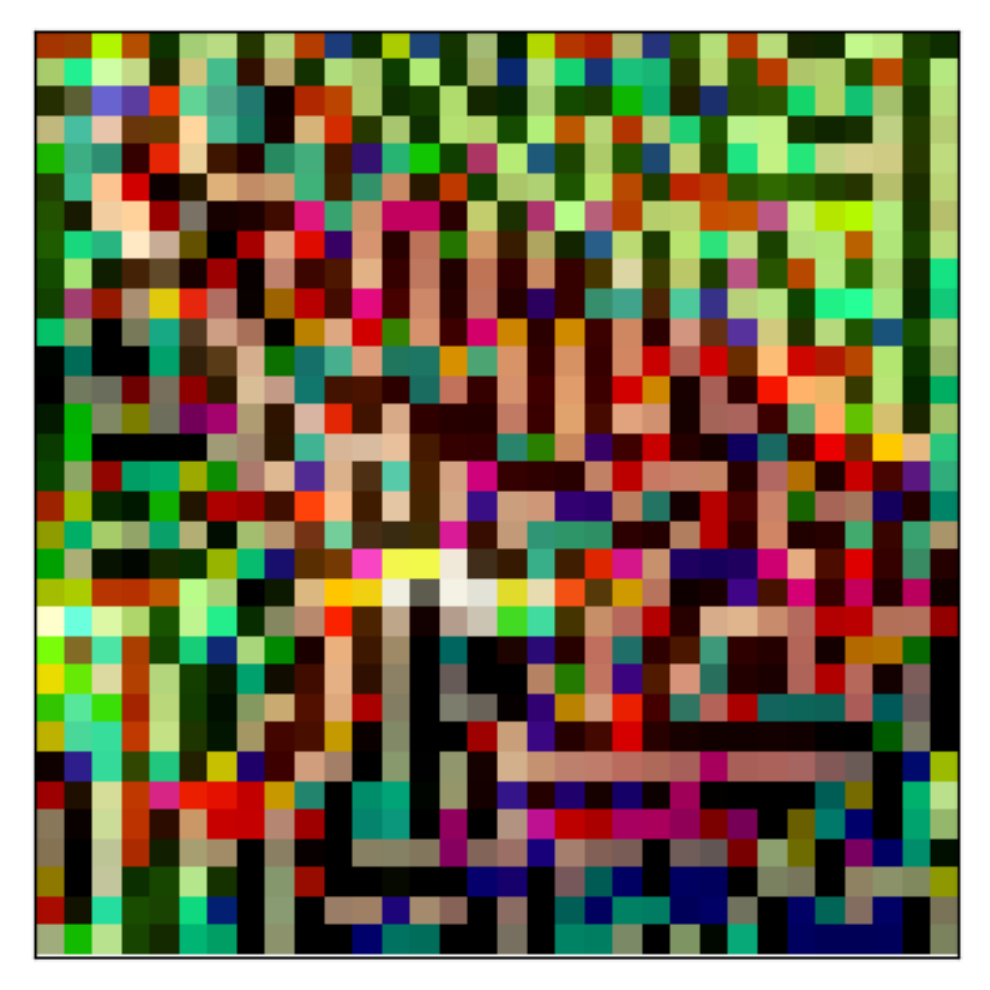}\\[-1ex]
		Frog($99.9\%$)& Frog($90.3\%$) & Frog($93.0\%$) & Frog($97.6\%$) & Frog($91.6\%$)\\
		\includegraphics[width=.18\linewidth]{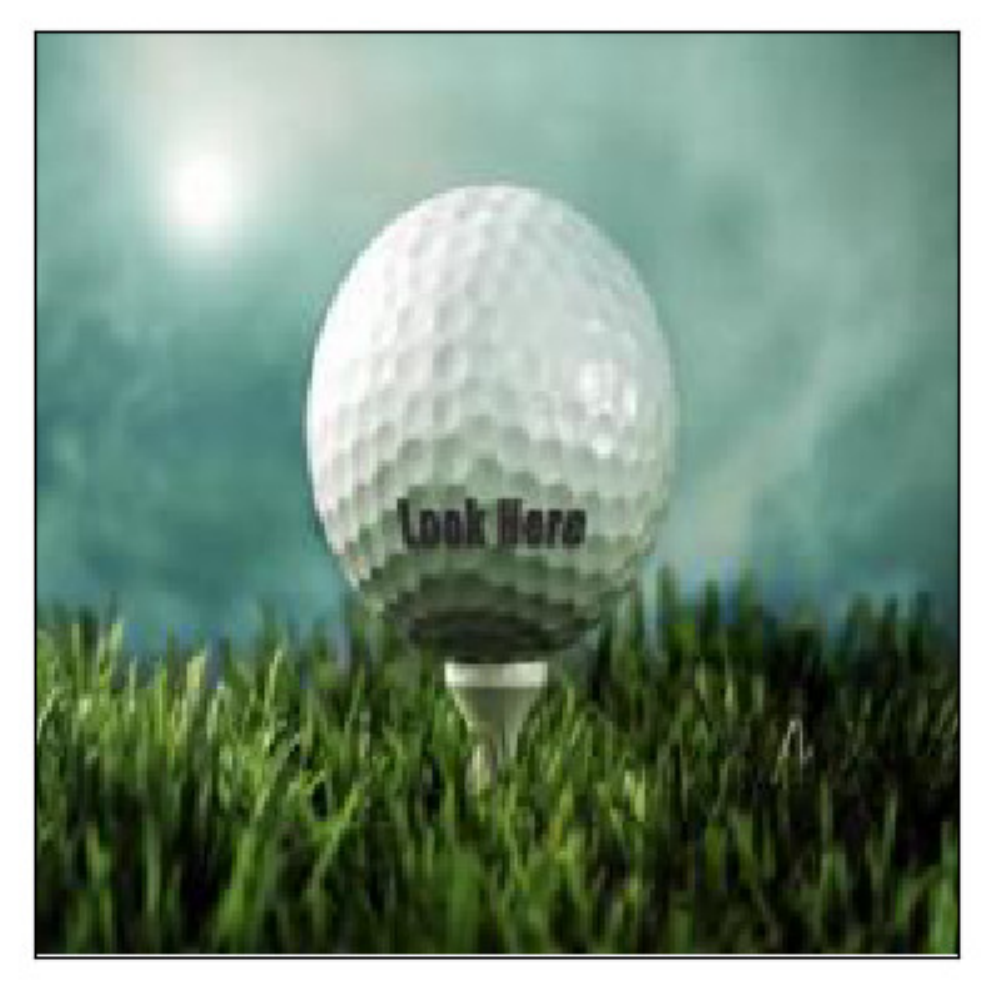}&
		\includegraphics[width=.18\linewidth]{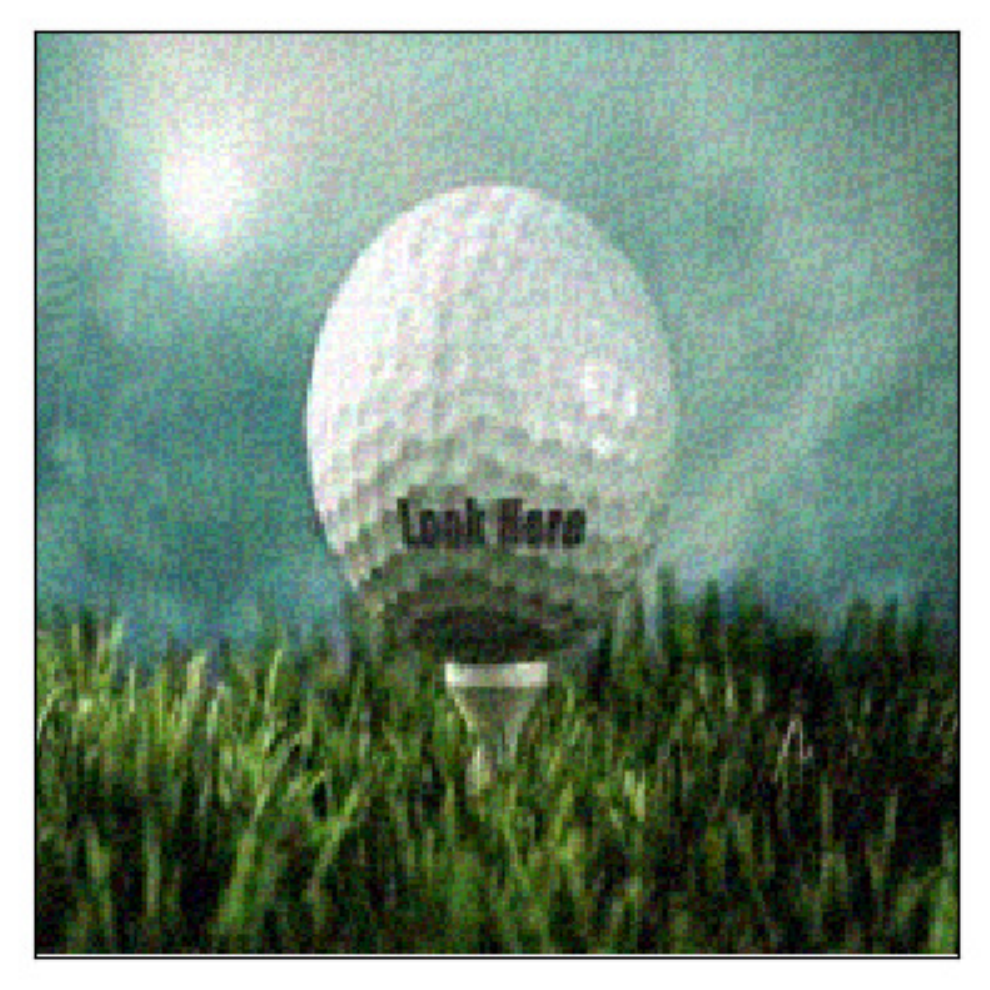}&
		\includegraphics[width=.18\linewidth]{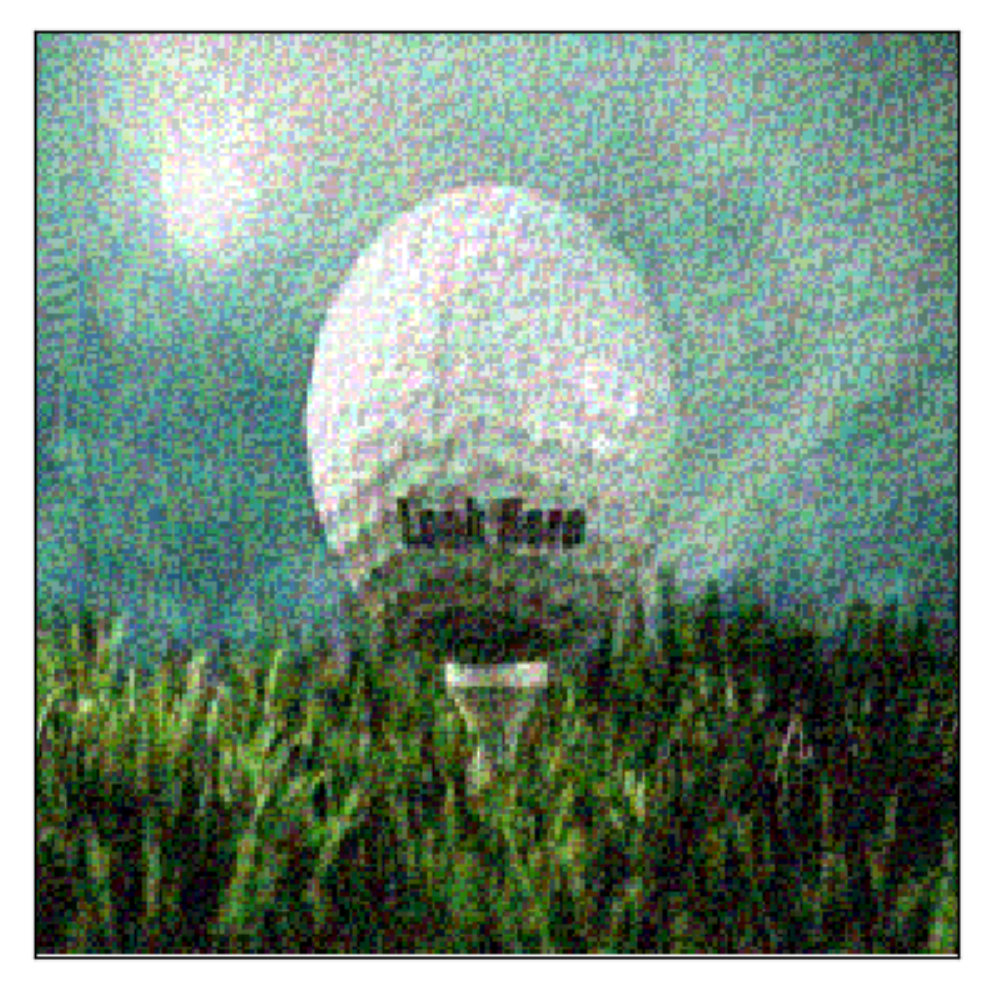}&
		\includegraphics[width=.18\linewidth]{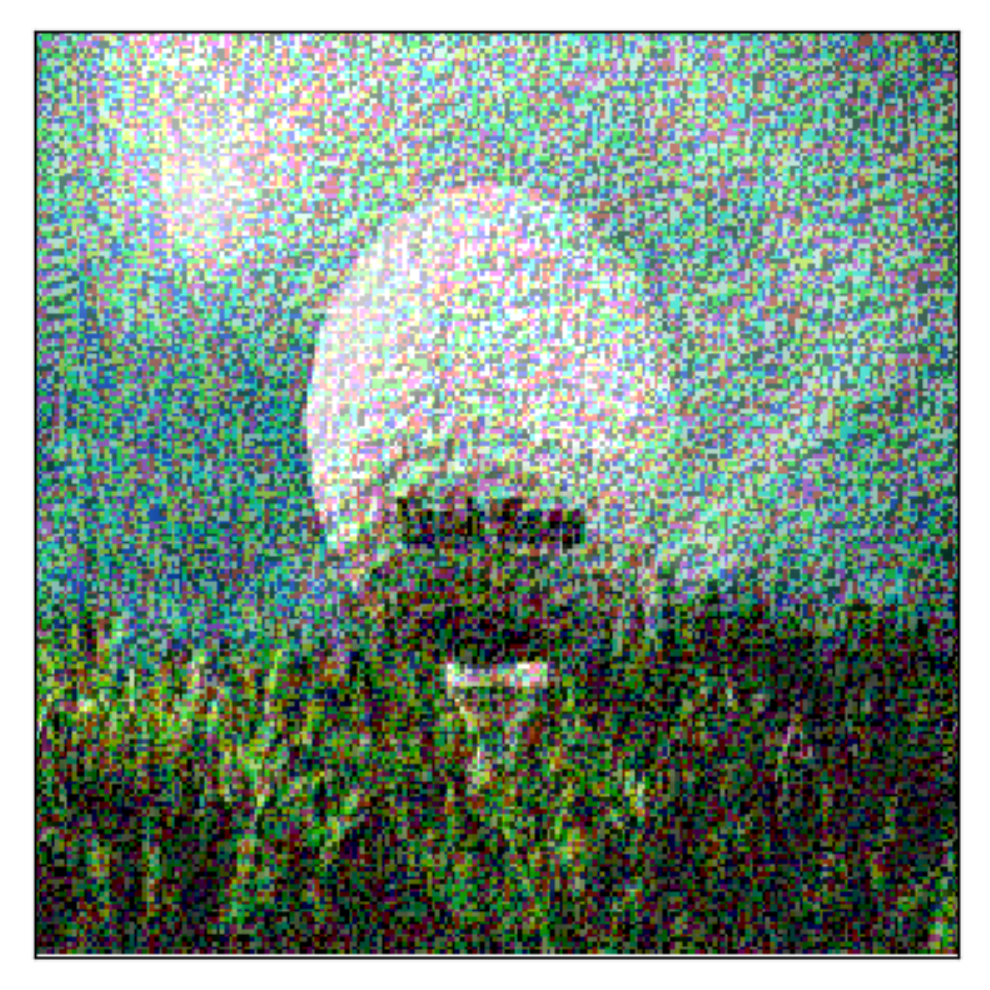}&
		\includegraphics[width=.18\linewidth]{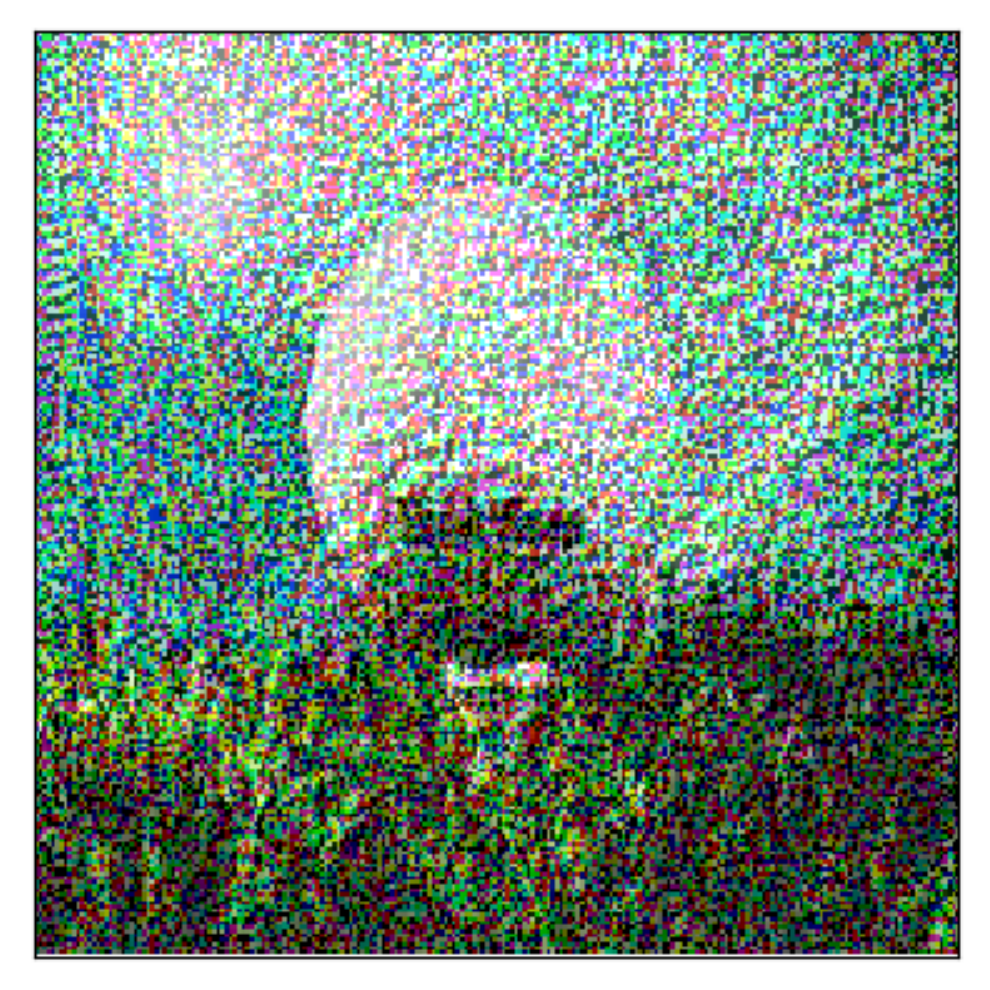}\\[-1ex]
		Ball($99.9\%$) &Ball($99.9\%$) &Ball($99.8\%$) &Ball($95.5\%$) &Ball($93.3\%$)\\
	\end{tabular}
	\caption{Use different perturbation sizes $\epsilon$ to attack the image. The first column is the original image, and the following columns are the images after the attacks using $\epsilon=$ 0.05, 0.1, 0.2, and 0.3.}%
	\label{fgsm_short}
\end{figure*}

The existing gradient-based attacks are mostly FGSM and its variants. Because the training of the DNN converges in the direction of loss reduction by updating the weights in the gradient decent direction, these methods attempt to increase the loss by introducing a slight perturbation in the gradient direction to make model suggest a wrong prediction. Further, these methods have indeed achieved obvious attack effects\cite{dong2018boosting}.

However, it must be noted that adding a perturbation in the gradient direction does not necessarily increase the loss and cause the model to predict errors. We can look at two examples attacked by FGSM, as shown in Fig. \ref{fgsm_short}. From left to right, as the size of the perturbation increases, the classifiers still have high confidence that can be correctly classified. When $\epsilon$ is 0.3, the added noise causes a serious deformation to the image; however, the classifier still has the correct classification with more than 90\% confidence. This shows that there are some cases in which a small enough (tolerant) disturbance in the gradient direction cannot cause classifier misclassification, which is possible in highly complex neural networks. Moreover, in the first row of Fig. \ref{fgsm_short}, when $\epsilon$ is 0.1, 0.2, and 0.3, the confidence classified as the category of frog is higher than the confidence when $\epsilon$ is 0.05. Therefore, in the gradient direction, as the amount of disturbance increases, the prediction confidence of the model for the correct category does not necessarily decrease. 

From the above phenomena, it is clear that the effective attack directions for the samples may not be near the gradient, and they may also be deviated greatly from the gradient direction. In this paper, RDA based on a hill climbing search is proposed for identifying an effective attack direction.

Hill climbing is an iterative local optimization algorithm to solve optimization problems. It starts from a random solution and attempts to find a better solution by moving the current solution to its better/best neighbor. If the better solution is found, it replaces the current solution. These steps are repeated until no better solution (neighbors) can be found. Based on the strategy of selecting the next solution, different types of hill climbing approaches can be employed: steepest hill climbing selects the best solution from the candidate neighbors; random hill climbing selects the candidate neighbors randomly; first-choice hill climbing of selecting the first random neighbors; and random restart hill climbing restarts the algorithm when the search fails\cite{russell2016artificial}.

\subsection{Random Directional Attack}
\subsubsection{Strategies}
\label{sec:Strategies}
The selection of the attack direction in RDA is based on first-choice hill climbing. Each iteration randomly seeks the first direction that can reduce the confidence of the attacked category by randomly rotating a small angle in the current direction. Of course, there may be other more advanced optimization algorithms for solving the optimal disturbance direction problem than hill climbing, such as simulated annealing\cite{kirkpatrick1983optimization} and evolutionary algorithms \cite{engelbrecht2007computational}. However, in a demanding real-time attack system, the relative simplicity of the hill climbing method makes it easier to converge to an effective attack. To reduce the time complexity and solve the problem of being trapped in a local optimum, our algorithm uses several strategies as follows.

\begin{itemize}
\item \textbf{The gradient direction is used as the initial direction}: Instead of randomly generating the initial point as the general hill climbing, we use the gradient direction as the initial solution. Although attacks in the gradient direction may not be able to successfully attack the model, the gradient-based methods such as FGSM provide inspiration in that a direction that may reduce the probability of the attacked category can be selected as a candidate when searching for the effective attack direction.
\item \textbf{Rotation on a random part of dimensions}: If all dimensions are rotated, the direction change will be large. That is, if the step size in the hill climbing is too long, it may skip the optimal value. Therefore, we choose a part of dimensions to rotate, and to enhance the diversity of the search, the rotation dimensions of each iteration are randomly selected.
\item \textbf{Variety of rotation angles}: The angle of rotation in each iteration of hill climbing is uniformly selected within a large range, which makes it possible for some dimensions to change significantly. This could increase the exploration ability of the algorithm and maintain the diversity of the searched solutions.
\end{itemize}

\subsubsection{Algorithm Framework}
\begin{algorithm}[t]
	\caption{RDA}
	\hspace*{0.02in}{\textbf{Input:}}
	The original image $x$, true class $y$, and the trained model $F(x,\mu)$ \\
	\hspace*{0.02in}{\textbf{Output:}}
	The adversarial example $x^*$
	\begin{algorithmic}[1]
		\State initialize $success=False$, $i=0$
		\State $\vec{v}_0=\bigtriangledown_xJ(\mu,x,y)$
		\State $cur\_best=F_y(x+\epsilon\cdot sign(\vec{v}_0))$
		\Repeat
		\State $R=Generate\_Rotation\_Matrix\_Set()$
		\label{GRMS}
		\State $shuffle(R)$
		\State $Setp\_Forward=false$
		\For{$j=1$ \textbf{to} $len(R)$}
		\State $P=F(x+\epsilon\cdot sign(R_j\cdot \vec{v}_i))$
		\label{rotate1}
		\If{$argmax(P)\ne y$ or $P_y<cur\_best$}
		\State $cur\_best=P_y$
		\State $success=(argmax(P)\ne y)$
		\State $\vec{v}_{i+1}=R_j\cdot \vec{v}_i$
		\label{rotate2}
		\State $i++$
		\State $Setp\_Forward=true$
		\State break
		\EndIf
		\EndFor
		\Until{$success==True$ or $Setp\_Forward==false$}
		\State \Return $x+\epsilon\cdot sign(\vec{v}_i)$
	\end{algorithmic} 
	\label{alg1} 
\end{algorithm}

Although some works show that an one-step attack of FGSM has a limited attack performance \cite{dong2018boosting,kurakin2016adversarial}, the RDA proposed in this paper still uses the one-step attack. 
We will prove that one-step attacks are feasible, and we will further improve the performance of such attacks. 
The method of generating an adversarial example by RDA is as follows:

\begin{equation}
x^*=x+\epsilon\cdot sign(\vec{v})
\label{equ:rda_x}
\end{equation}

This formula is similar to that used in FGSM; however, it is important to note that $\vec{v}$ here indicates the direction of the successful attack, which is not necessarily the direction of gradient. We use first-choice hill climbing and combine the three strategies described above to obtain the specific direction to achieve a successful attack. We call the attack method based on random but not arbitrary direction as a random directional attack, and the specific procedure is shown in Alg. \ref{alg1}.

First, the gradient of the model for the input $x$ is calculated and set as the initial direction.
Let $cur\_best$ denote the confidence that the sample disturbed is classified into the true category $y$. 
The initial $cur\_best$ is set to the confidence that the sample disturbed at the gradient direction is classified into the true category $y$ (we select the confidence of category $y$ as the evaluation index, and search the attack direction by minimizing this value). 

Then, in each subsequent iteration, we construct the rotation matrix set and shuffle it.
The first rotation matrix that made the $cur\_best$ smaller is selected, and it is used to update $cur\_best$ and the attack direction.
If the direction at the end of an iteration can cause a misclassification, the attack succeeds and the algorithm ends. 
If all rotation matrices cannot make $cur\_best$ smaller, the attack fails and the algorithm ends.

It should be noted that, according to (\ref{equ:rda_x}), although RDA uses multiple iterations to determine the attack direction, the attack is essentially a one-step attack. In addition, under white-box attacks, RDA takes the gradient direction of the target model as the initial direction; However, under black-box attacks, RDA takes the gradient direction of the substitute model as the initial direction.

\begin{algorithm}[t!]
	\caption{Generate Rotation Matrix Set}
	\hspace*{0.02in}{\textbf{Input:}}
	The number of dimensions $m$, number of selected dimensions $l$ ($l$ is an even number), and rotation angle range $[-\theta,\theta]$\\
	\hspace*{0.02in}{\textbf{Output:}}
	The rotation matrix set $R$
	\begin{algorithmic}[1]
		\State initialize set $R=emptyset()$
		\For{$\beta=-\theta$ \textbf{to} $\theta$} //except $\beta=0$
		\State randomly select $l$ dimensions from $m$ dimensions to 
			\par \setlength{\parindent}{1.5em} compose a vector $v=[v_1,v_2,...,v_l]$
		\State initialize $r$ to an $m$-dimensional unit matrix
		\For{$i=0$ \textbf{to} $l/2-1$}
		\State $r_{v[2i],v[2i+1]}=\left[\begin{matrix}
		cos(\beta) & sin(\beta)\\
		-sin(\beta) & cos(\beta) \\
		\end{matrix}\right]$
		\label{twodimentions}
		\EndFor
		\State $R=R\cup\{r\}$
		\EndFor
		\State \Return $R$
	\end{algorithmic} 
	\label{alg2} 
\end{algorithm}

\subsubsection{Generating a Rotation Matrix Set}
The procedure for generating a rotation matrix set in line \ref{GRMS} of Alg. \ref{alg1} is shown in Alg. \ref{alg2}.

For a specific angle $\beta$, on the basis of the identity matrix, we rotate two dimensions $r_{v[i],v[i+1]}$ in turn, as shown in line \ref{twodimentions} of Alg. \ref{alg2}. After all selected $l$ dimensions are rotated, the rotation matrix is obtained.

In the experiments, owing to the particularity of the rotation matrix, we do not even need to generate a specific rotation matrix set for each iteration. 
In addition, we do not need to store the entire rotation matrix set. 
We only need to calculate $r_{v[0],v[1]}$ for each angle $\beta$, and only store $r_{v[0],v[1]}$ as well as the selected $l$ dimensions for each angle $\beta$.
Furthermore, at lines \ref{rotate1} and \ref{rotate2} in Alg. \ref{alg1}, there is no need to perform complex matrix multiplication, and only a simple rotation of the selected $l$ dimensions is required. 
The number of multiplications for calculating each of the two dimensions is 4, and the total number of multiplications is $l/2\times4=2l$, which is independent of the dimensions ($m$) of the direction vector. 
Thus,  matrix preservation and matrix multiplication required for the traditional angle rotation are inexpensive in RDA.

Now, we consider the similarity between the current direction and rotated direction vectors.
Although the selected dimensions may rotate at a larger angle according to strategies mentioned in section \ref{sec:Strategies} (such as greater than 90 degrees or even 180 degrees), from the viewpoint of the entire direction vector, it does not violate the basic hill climbing strategy of selecting the neighbors (small step size; here, the rotation angle). Let the direction vectors before and after the rotation be $A$ and $B$, respectively. If the similarity of $A$ and $B$ is measured by cosine similarity, then,

\[
\begin{split}
similarity(A,B)&=\frac{A\cdot B}{\left\lVert A\right\lVert\left\lVert B\right\lVert}\\
&=\frac{\sum_{i=1}^{m}A_i\times B_i}{\sqrt{\sum_{i=1}^{m}(A_i)^2}\times \sqrt{\sum_{i=1}^{m}(B_i)^2}}.
\end{split}
\]

Let the number of selected dimensions be $l$. Without loss of generality, we assume that the first $l$ dimensions are selected. Considering that the lengths of the two vectors before and after the rotation are equal, we have

\[
\begin{split}
similarity(A,B)&=\frac{\sum_{i=1}^{l}A_i\times B_i+\sum_{i=l+1}^{m}(A_i)^2}{\sum_{i=1}^{m}(A_i)^2}\\
&=1-\frac{\sum_{i=1}^{l}A_i\times (A_i-B_i)}{\sum_{i=1}^{m}(A_i)^2}\\
&\ge 1-\frac{\sum_{i=1}^{l}2(A_i)^2}{\sum_{i=1}^{m}(A_i)^2}\\
&\approx 1-\frac{2l}{m}
\end{split}.
\]

When $l$ is 10, the similarity will be no less than 0.993 for a CIFAR-10 image of size $32\times32\times3$, and not less than 0.9999 for an ImageNet image of size $224\times224\times3$. The minimum value is obtained by rotating 180 degrees, and therefore, the extreme rotation of 180 degrees produces only a slight change to the entire direction vector. Therefore, the basic search strategy of selecting neighbors is not violated.

\subsubsection{Analysis}

\begin{figure}
	\centering
	\includegraphics[width=80mm]{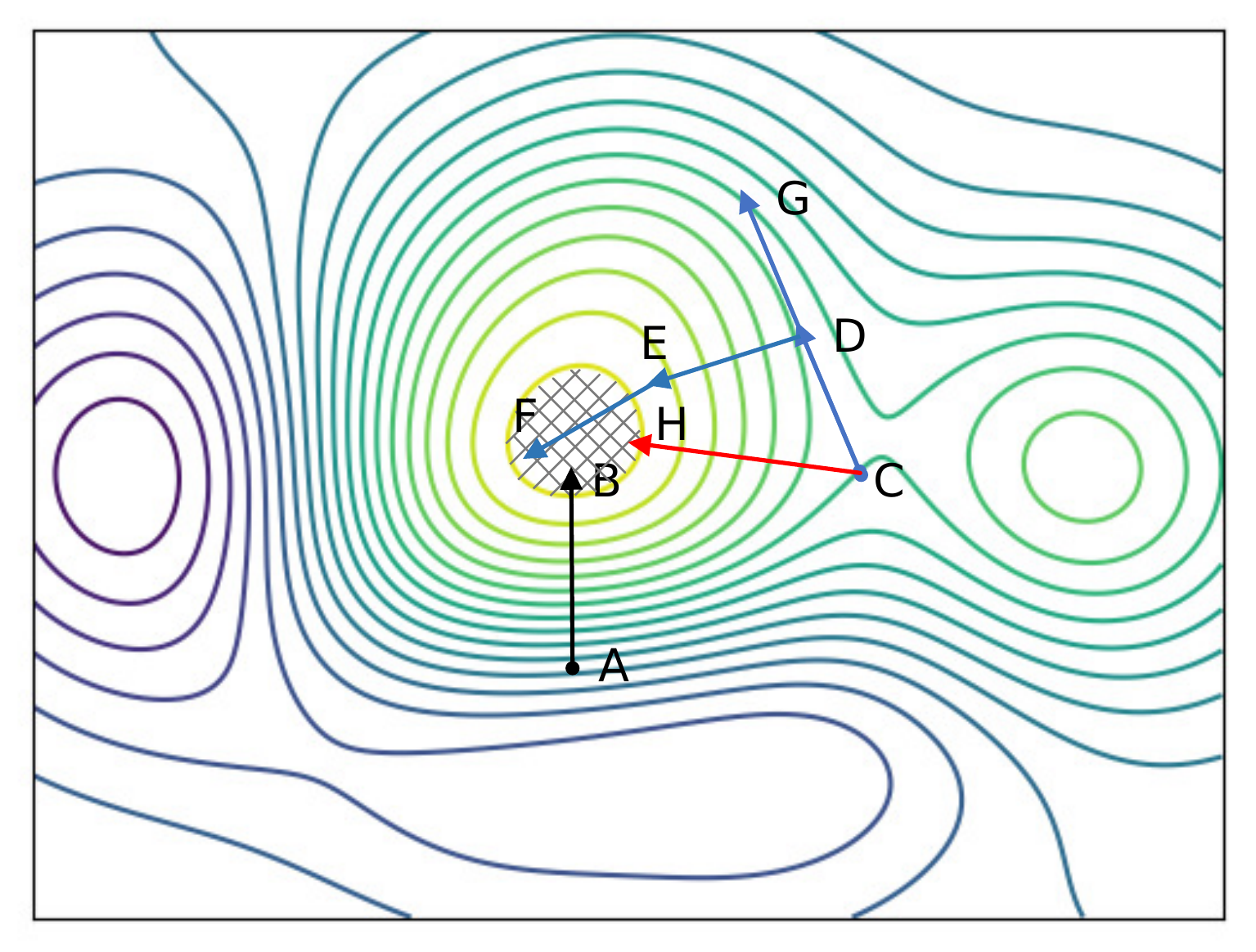}\\
	\caption{The simplified contour map of loss.}
	\label{loss}
\end{figure}

To more visually analyze RDA and the related attacks, we illustrate the directions of such attacks in Fig. \ref{loss}. 
This can be seen as a contour map of loss (loss in high-dimensional space is often very complex; here, it is expressed as a simple contour map for the ease of analysis), where the loss of points on each line is the same. 

The gradient direction of each point on the image will be in a direction vertical to the contour. Further, if the loss function of the neural network adopts cross entropy, i.e.,

$$loss=-\sum_{i}p_i(x) log(F_i(x)),$$
where $p_i(x)$ represents the probability that the classifier classifies input $x$ as category $i$, for the single-label classifier that uses the one-hot encoded label vector, $loss=-log(F_y(x))$. Therefore, Fig. \ref{loss} could be used to represent the classification confidence contour map of the true category.

First, we consider the point $A$ in Fig. \ref{loss}. The direction $\overrightarrow{AB}$ represents the gradient direction of point $A$. 
We assume that the shadow in the figure is the adversarial region, and the point reaching the area can cause misclassification (note that the adversarial region is complex, and the increasing loss only represents that the adversarial example is more likely to be generated; this is not absolute. We have simplified this here). 
Therefore, if a perturbation such as the size of $|\overrightarrow{AB}|$ is added along the gradient direction $\overrightarrow{AB}$ at point $A$ to reach point $B$ in the adversarial area, an adversarial example can be generated. 
This is the main idea of the attack of FGSM.

However, the high complexity of DNNs also makes the case not always true. If the input $x$ is at point $C$, the gradient direction is along the direction of $\overrightarrow{CD}$. It can be observed that when the perturbation is added along the gradient direction at point $C$, the loss after adding a perturbation of size $|\overrightarrow{CG}|$ is smaller than the loss of adding a smaller perturbation $|\overrightarrow{CD}|$. 
This explains the phenomenon in Fig. \ref{fgsm_short} that the predicted probability does not decrease as the size of the perturbation increases. Moreover, regardless of how much perturbation is added to the direction $\overrightarrow{CD}$, it cannot be moved to the adversarial area, which is the limitation of the FGSM. Therefore, a variety of iterative-based gradient direction attacks are derived, which approach the adversarial area by multiple movements toward the gradient direction, such as $C\to D\to E\to F$ in Fig. \ref{loss}. Of course, they achieved a better attack effect.

These iterative-based methods lead us to think as follows: should we use multiple steps to reach the adversarial region? 
Let us consider point $C$ again. If we do not base the gradient direction and add a perturbation of size $|\overrightarrow{CH}|$ along the direction of $\overrightarrow{CH}$, point $C$ moves to point $H$ in the adversarial region, producing the same effect as the iterative version of FGSM. 
The random directional attack we proposed could make the attack direction move from the gradient direction $\overrightarrow{CD}$ to $\overrightarrow{CH}$ steadily through the idea of the hill climbing together with three strategies in section \ref{sec:Strategies}; finally, we realize the one-step attack successfully. In the next section, we further demonstrate the effectiveness of RDA attack by the experiment.

\section{Experiment and Evaluation}
\label{sec:experiment}
In this section, first, we introduce the experimental settings of the dataset and models training, and then comprehensively test the attack performance of RDA. Finally, we compare the proposed RDA with the typical gradient-based attacks.

\subsection{Experimental Setup}
Our experiments are based on cleverhans\cite{papernot2018cleverhans}, which is a standardized implementation of most current classical attack methods, including FGSM, L.L.Class, BIM, and MI-FGSM. Our source code is available at \url{https://github.com/Daftstone/Random-Directional-Attack}.

\subsubsection{Datasets}
To verify the reliability and versatility of our algorithm, four commonly used datasets, i.e., MNIST, SVHN, CIFAR-10, and ImageNet, were used in the experiments.

The MNIST and SVHN datasets contain images with 10 numbers from 0-9. The MNIST consists of a grayscale image of size $28\times28$, and the SVHN consists of color images of size $32\times32$. SVHN is a natural image set extracted from the house number photographed by Google Street View, and therefore, the processing of this dataset is more complicated and difficult.
CIFAR-10 contains $32\times32$ color images of 10 mutually exclusive categories such as airplanes, trucks, dogs, and frogs. To verify the attack performance of RDA on high resolution images, we chose the ImageNet dataset. 
We extracted a subset of 10 categories (ImageNet-10) for testing. The selection of these 10 categories is consistent with Imagenette (\url{https://github.com/fastai/imagenette}), including golf, parachute, and tent, and all images are resized to $224\times224\times3$; the average number of samples per category is about 1000.

\subsubsection{Model Training}

\begin{table}[htbp]
	\centering
	\caption{Classification accuracy (\%)}
	\begin{tabular}{l|rrrr}
		\hline
		model & \multicolumn{1}{l}{MNIST} & \multicolumn{1}{l}{SVHN} & \multicolumn{1}{l}{CIFAR-10} & \multicolumn{1}{l}{ImageNet-10} \\
		\hline
		Target & 99.24 & 96.29 & 91.37 & 91.80 \\

		Substitute  & 99.33 & 94.64 & 89.88 & 89.40 \\
		\hline
	\end{tabular}%
	\label{classification_acc}%
\end{table}%

We tested the white-box and black-box attacks separately, and therefore, the target model to be attacked and the substitute model for the black-box attack are required for each dataset. 

For MNIST, we use a simple 7-layer CNN structure as the target model and another 8-layer CNN model as a substitute model. For the SVHN dataset, both the target model and the substitute model use a 7-layer CNN structure; however, the structure used is different. CIFAR-10 uses a simple and practical VGG-16 as the target model and ResNet-32 as the substitute model. ImageNet-10 uses a more advanced lightweight network MobileNet and MobileNetV2 as the target and substitute models, respectively. 

The training process for the four datasets is the same. First, the images are normalized to [0, 1], and the initial learning rate is set to 0.001. If the accuracy of the test set is not improved, the current learning rate is reduced by 70\%. We use the Adam optimizer and data augmentation operations including rotation, horizontal vertical shift, and horizontal flip. All datasets are trained for 100 epochs. At the end of the training, a good classification accuracy was achieved in the four datasets, as shown in Table \ref{classification_acc}.

For all datasets, the samples misclassified by the models are not considered when we run the algorithms to generate the adversarial examples.

\subsection{Experimental Results}
In this section, we use white-box attacks for empirically analyses.

\subsubsection{Number of Selected Dimensions}

\begin{figure*}
	\centering
	\includegraphics[width=0.48\linewidth]{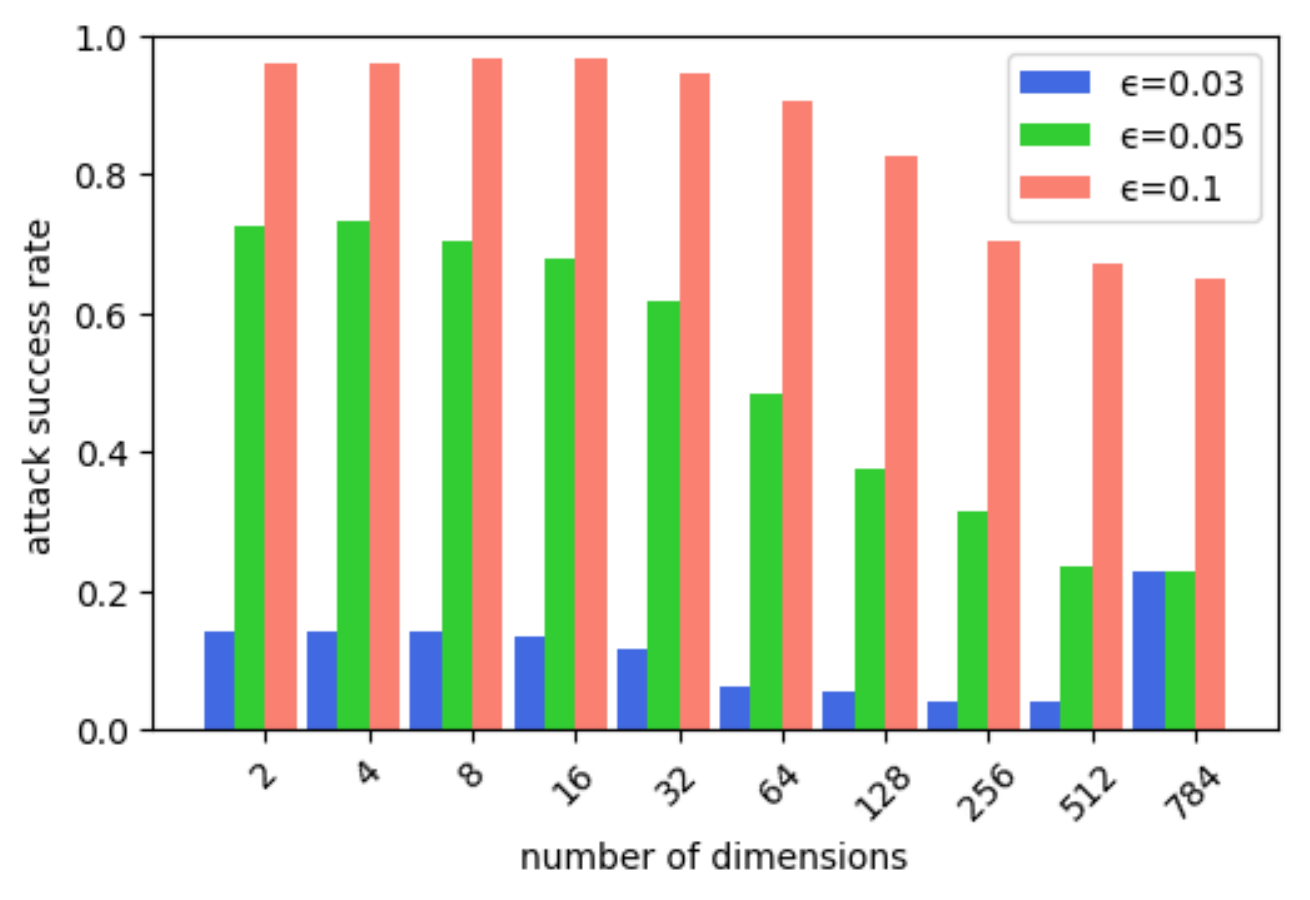}\vspace{4pt}
	\includegraphics[width=0.48\linewidth]{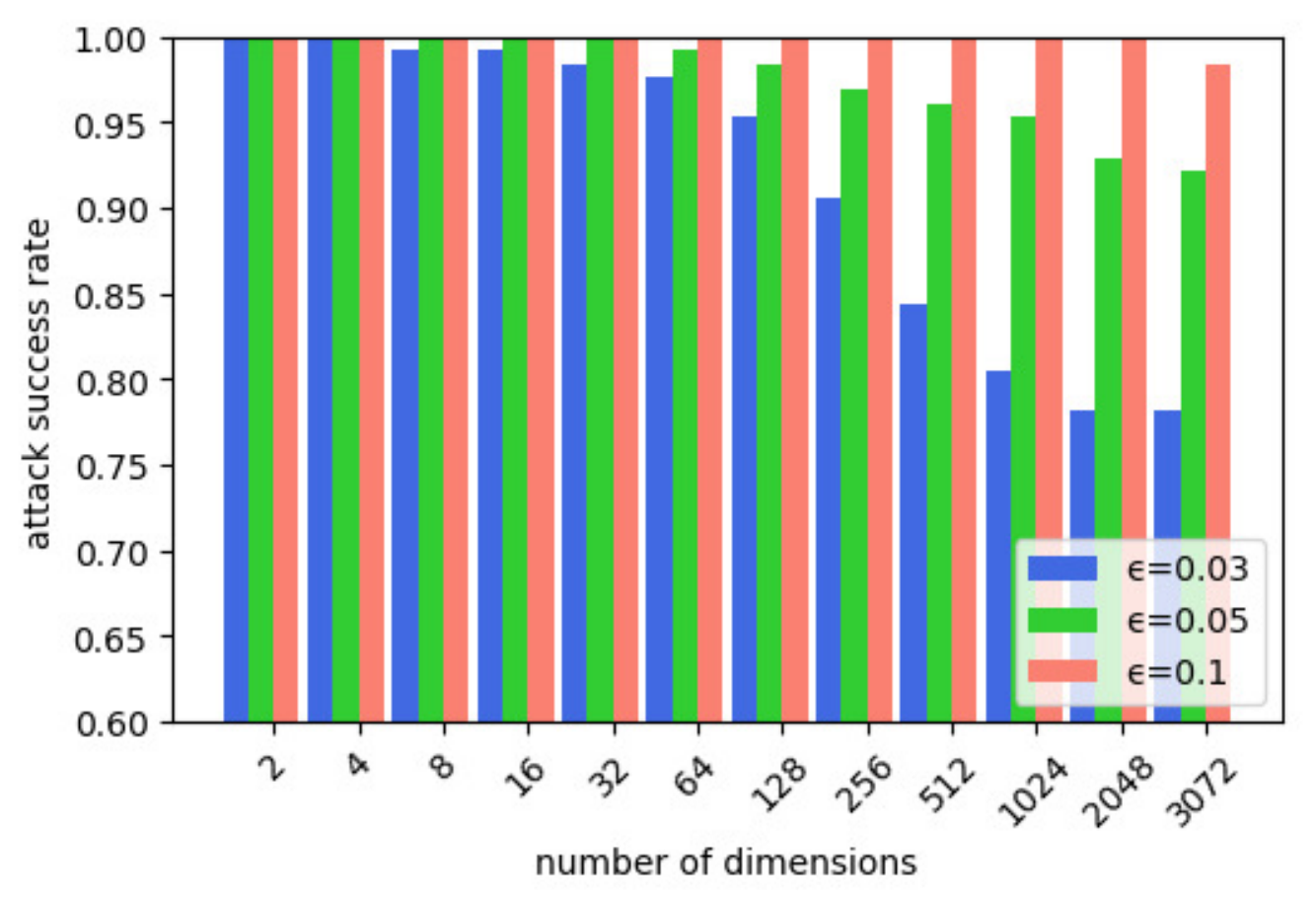}\vspace{4pt}
	\includegraphics[width=0.48\linewidth]{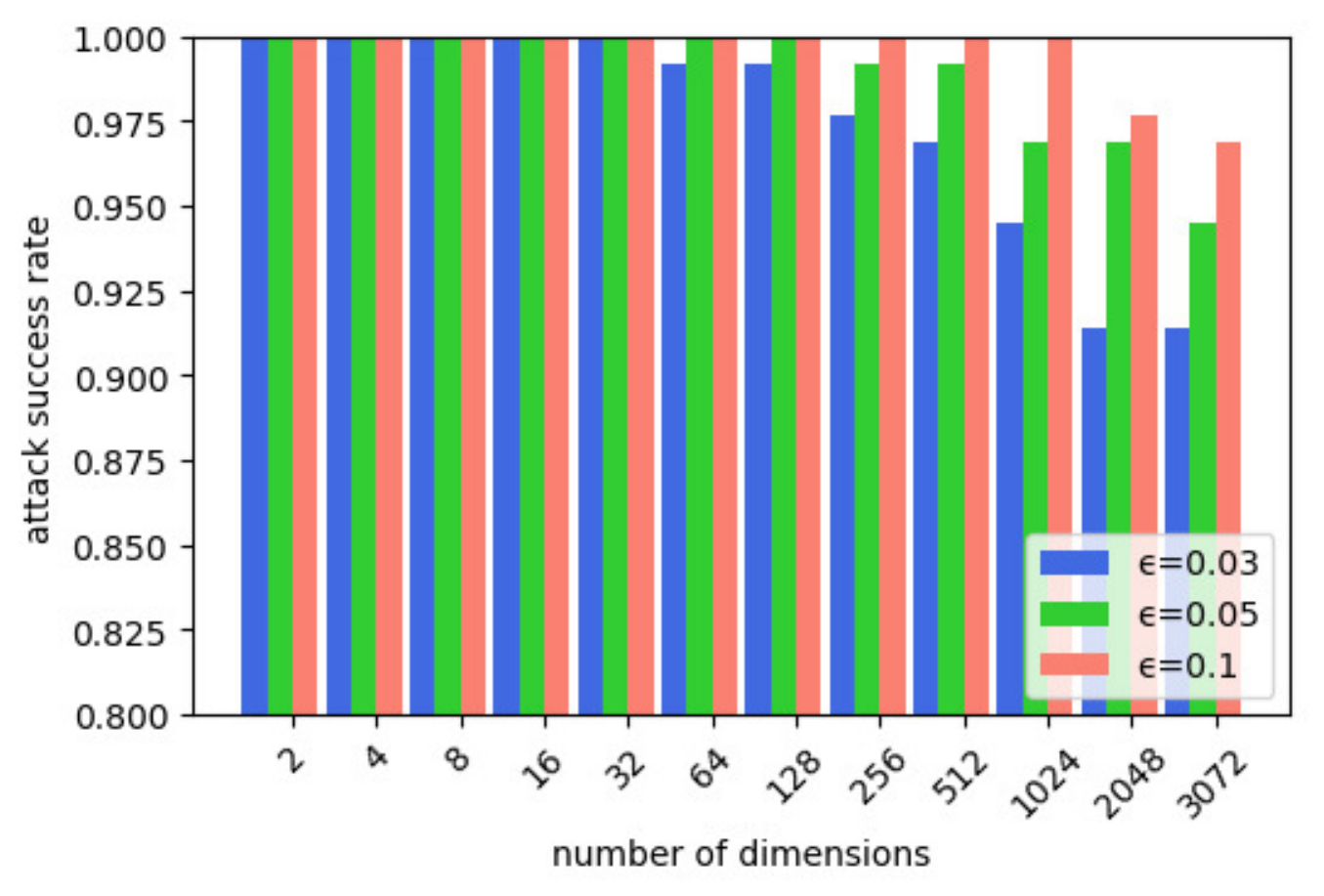}\vspace{4pt}
	\includegraphics[width=0.48\linewidth]{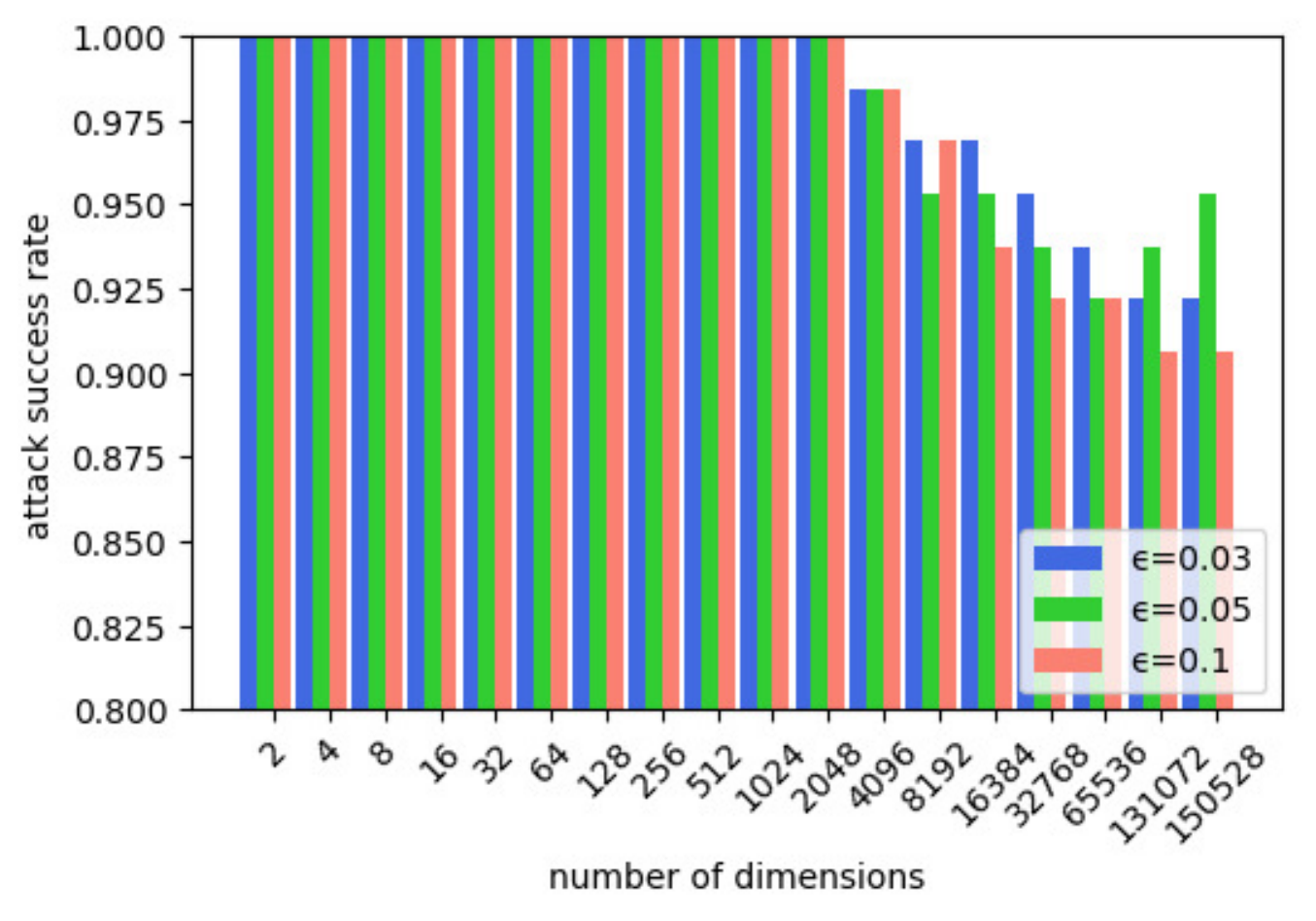}
	\caption{The effect of the number of selected dimensions $l$ on the attack performance. The results of testing on MNIST, SVHN, CIFAR-10 and ImageNet-10 are from left to right and then top to bottom.}
	\label{dimensions}
\end{figure*}

RDA uses randomly selected dimensions to ensure the diversity of the search and to cover as many directions as possible. 
This section examines the effect of the number of selected dimensions $l$ on the success rate of attacks. We set $l$ to the $n$th power of 2 and all dimensions $m$ in the experiments, and the angle of rotation is 180 degrees. 
The experimental results obtained by calculating the attack success rate when $\epsilon$ is 0.03, 0.05, and 0.1 are shown in Fig. \ref{dimensions}.

As can be seen from Fig. \ref{dimensions}, in each dataset, as the number of dimensions increases from left to right, the attack success rate decreases, and in most cases, the worst attack performance occurs when selecting to rotate all dimensions. This also justifies the rationality of this paper using a part of the dimensions for rotation. Choosing more dimensions may not help identify the feasible direction, i.e., the hill climbing search may skip over the feasible directions. 

\begin{figure*}
	\centering
	\includegraphics[width=0.24\linewidth]{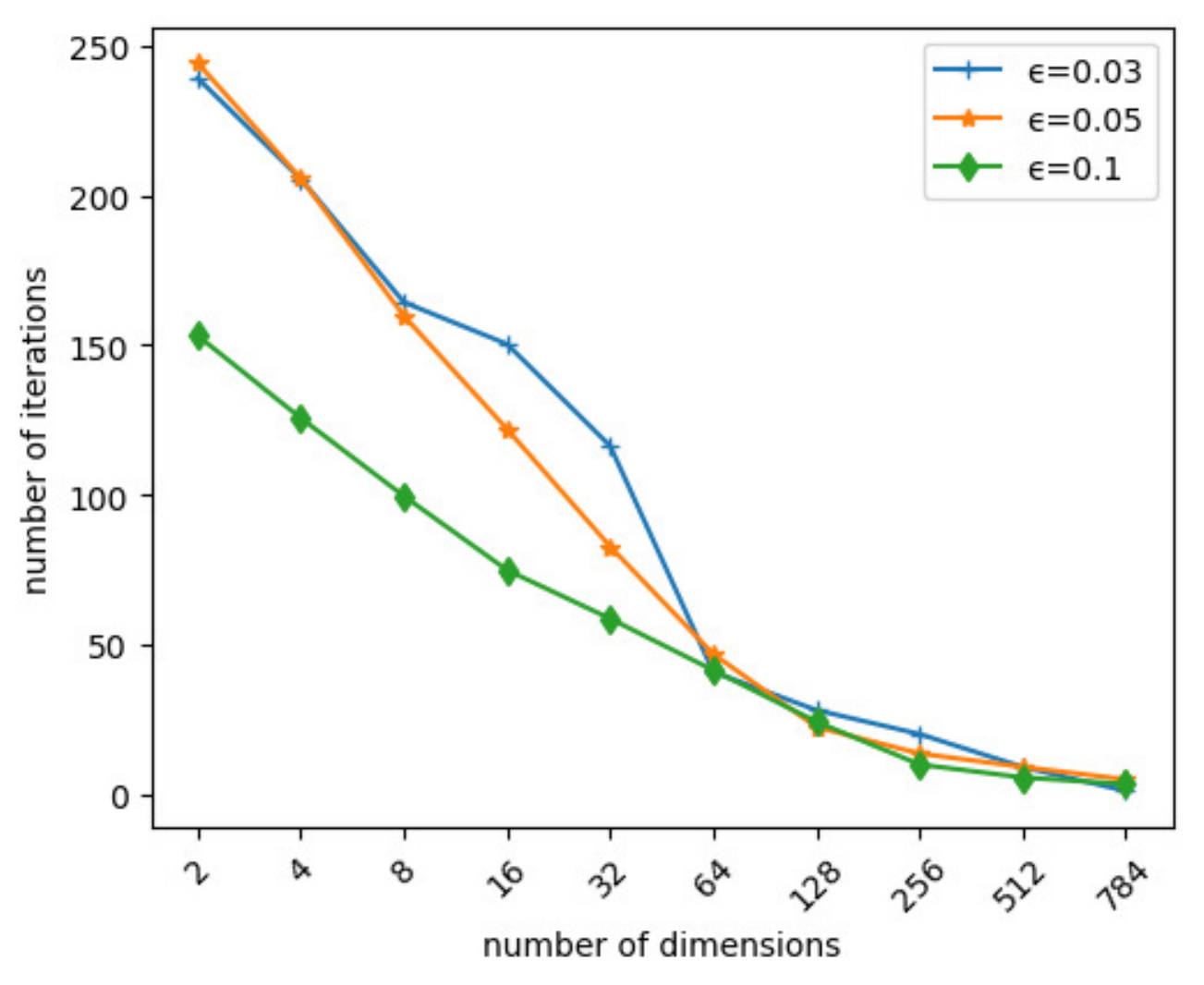}\vspace{4pt}
	\includegraphics[width=0.24\linewidth]{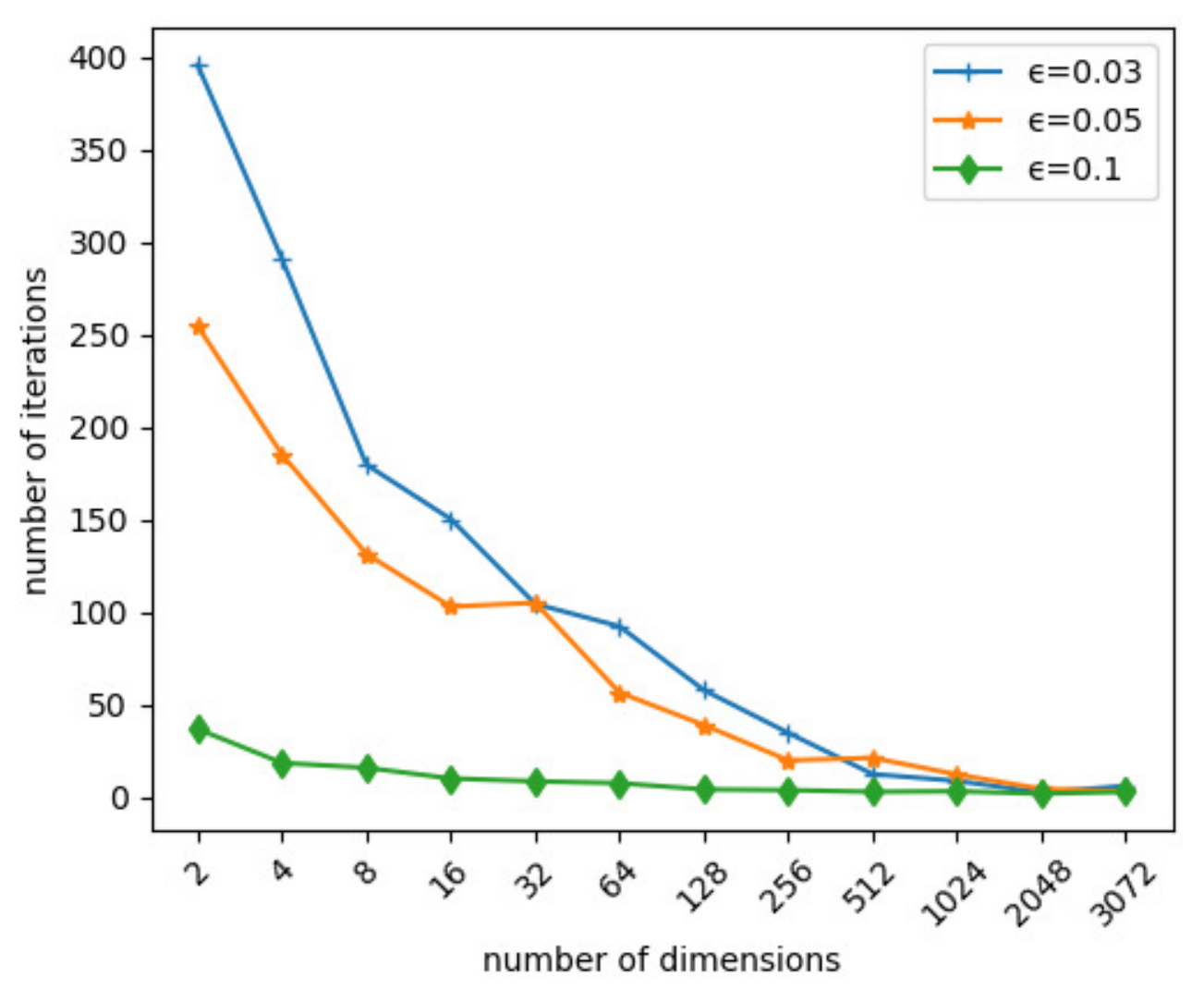}\vspace{4pt}
	\includegraphics[width=0.24\linewidth]{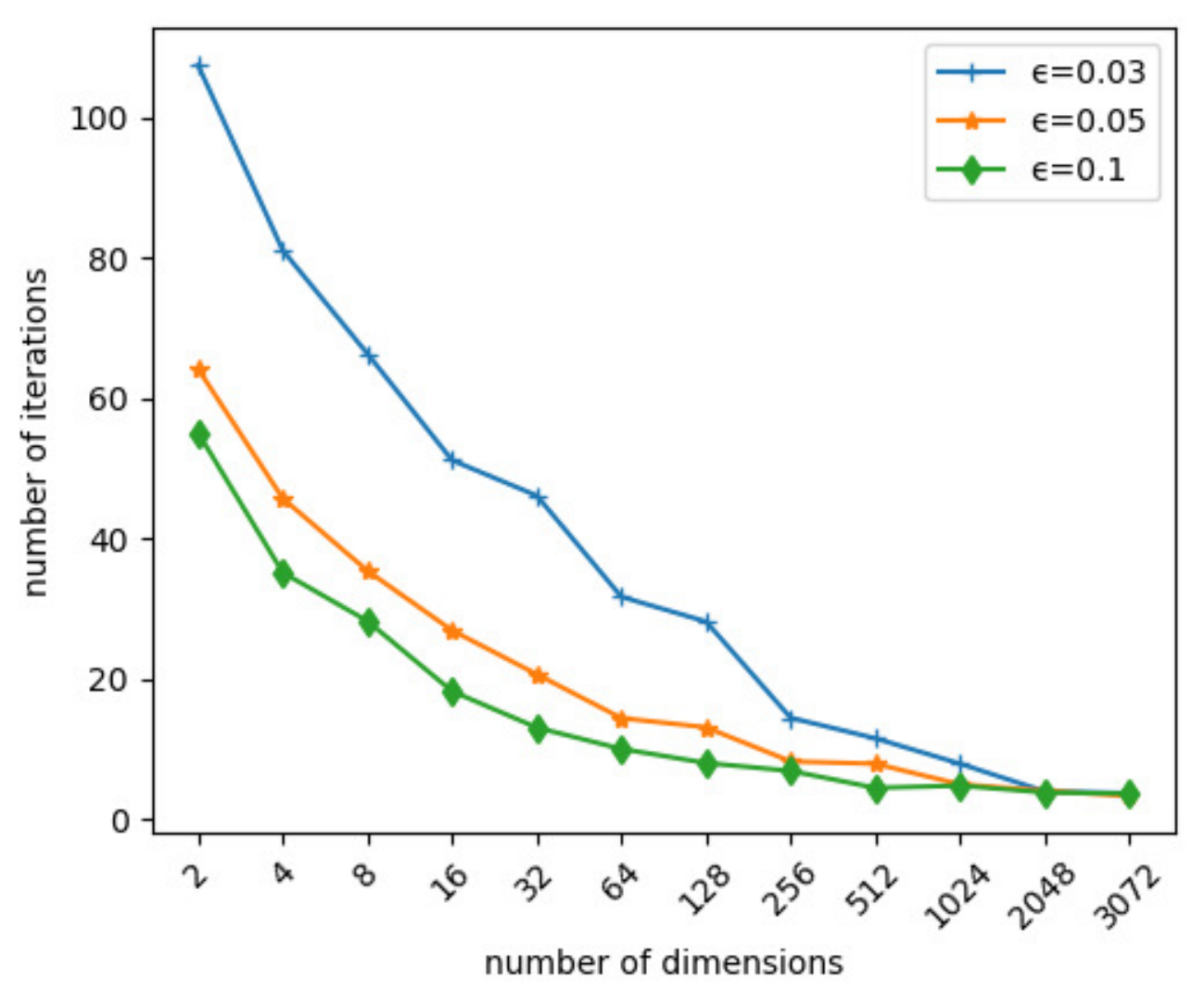}\vspace{4pt}
	\includegraphics[width=0.24\linewidth]{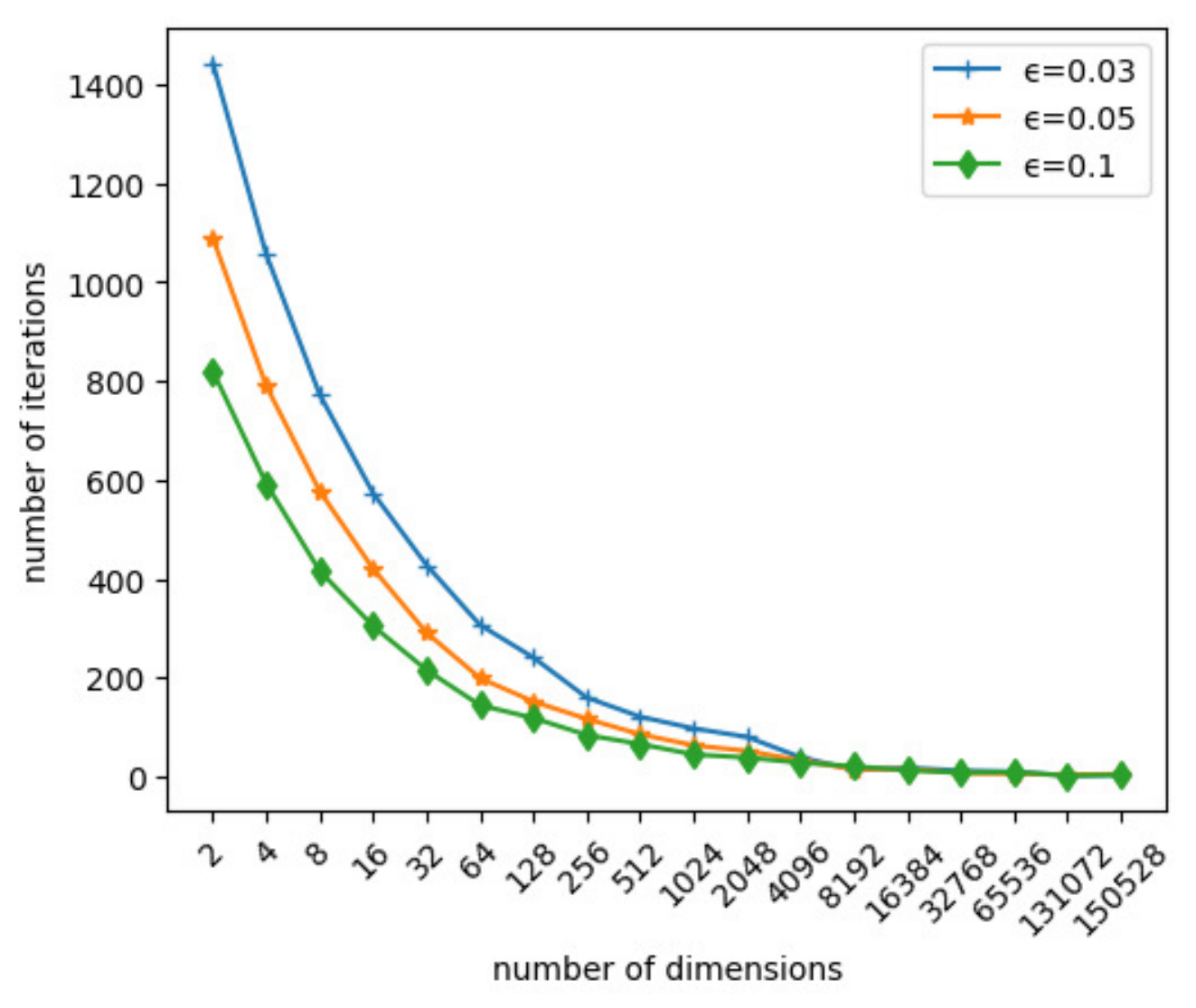}
	\caption{The relationship between the number of selected dimensions $l$ and the average number of iterations required by the algorithm. The results on the four datasets MNIST, SVHN, CIFAR-10 and ImageNet-10 are shown from left to right.}
	\label{dimension_iteration}
\end{figure*}

Although choosing a very small number of dimensions can help obtain better attack performance, the number of iterations that RDA needs to search for the effective direction might be larger. 
We counted the average number of iterations used to find the effective attack directions for the samples that could not be successfully attacked in the gradient direction, as shown in Fig. \ref{dimension_iteration}. 
As expected, the number of iterations is relatively large when less dimensions are selected. 
Further, as the number of dimensions increases, the number of iterations is reduced significantly.
We recommend that the number of selected dimensions is better when it is between 10 and 100.

\subsubsection{Angles of Rotation}

\begin{figure*}
	\centering
	\includegraphics[width=0.24\linewidth]{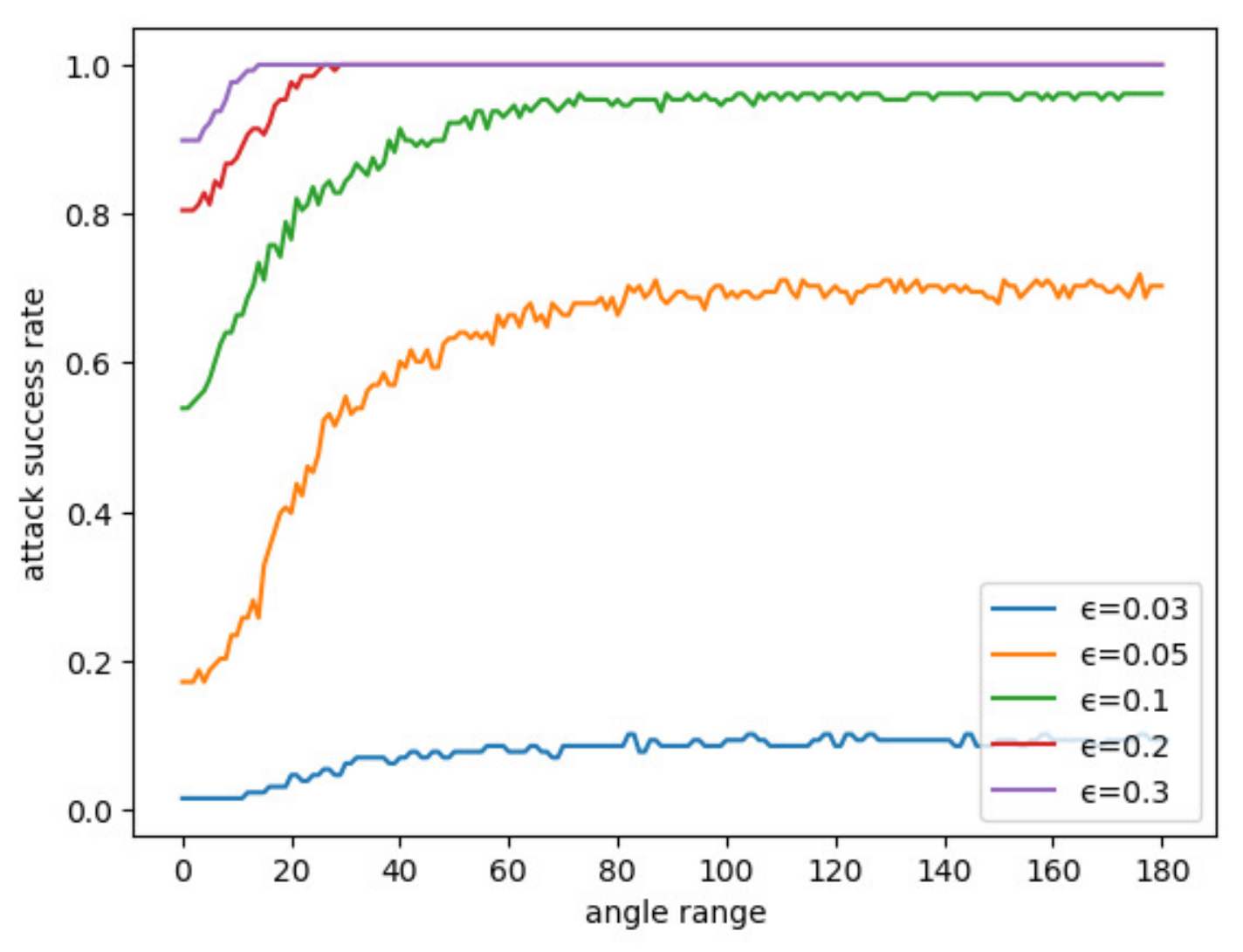}\vspace{4pt}
	\includegraphics[width=0.24\linewidth]{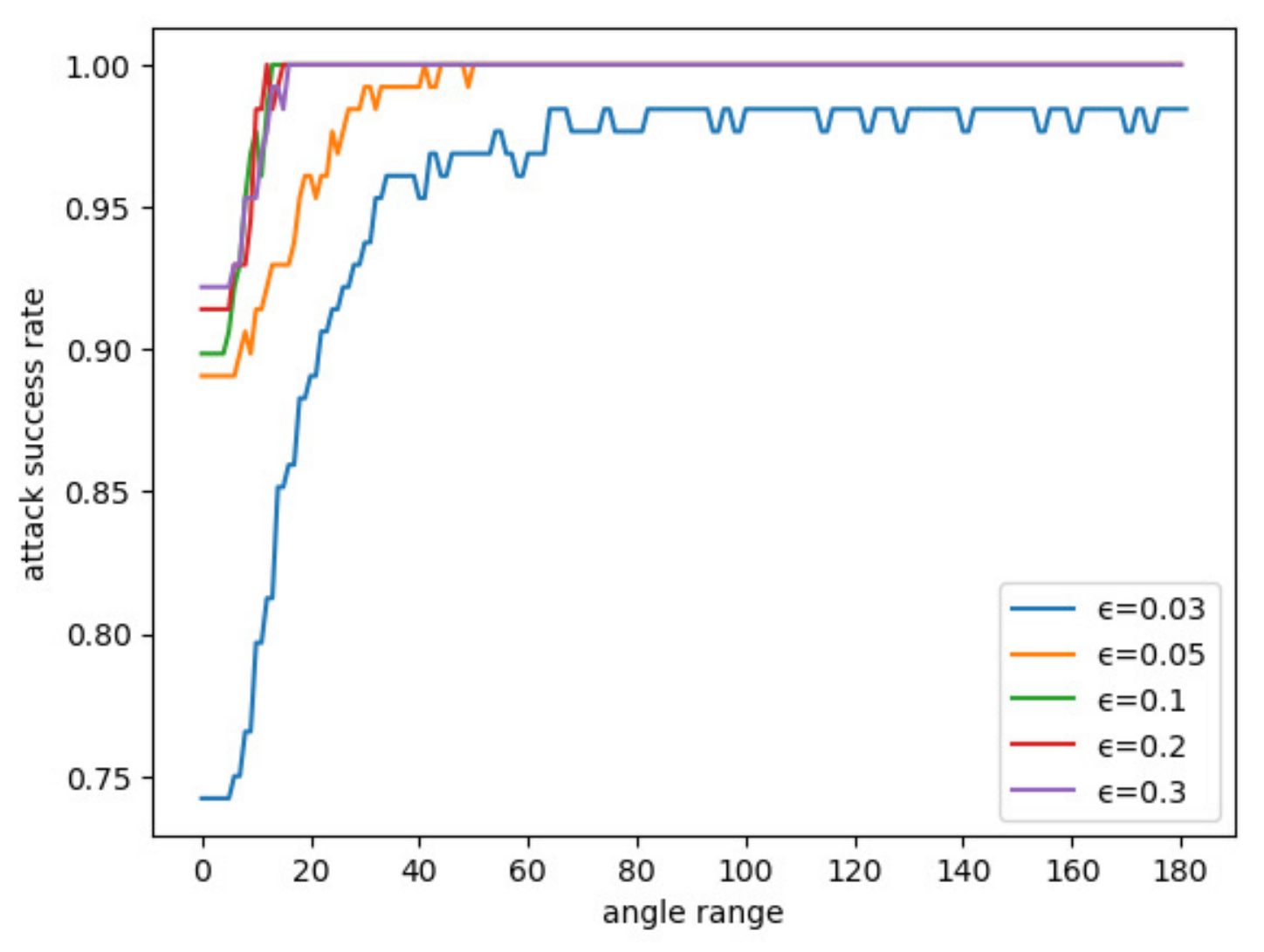}\vspace{4pt}
	\includegraphics[width=0.24\linewidth]{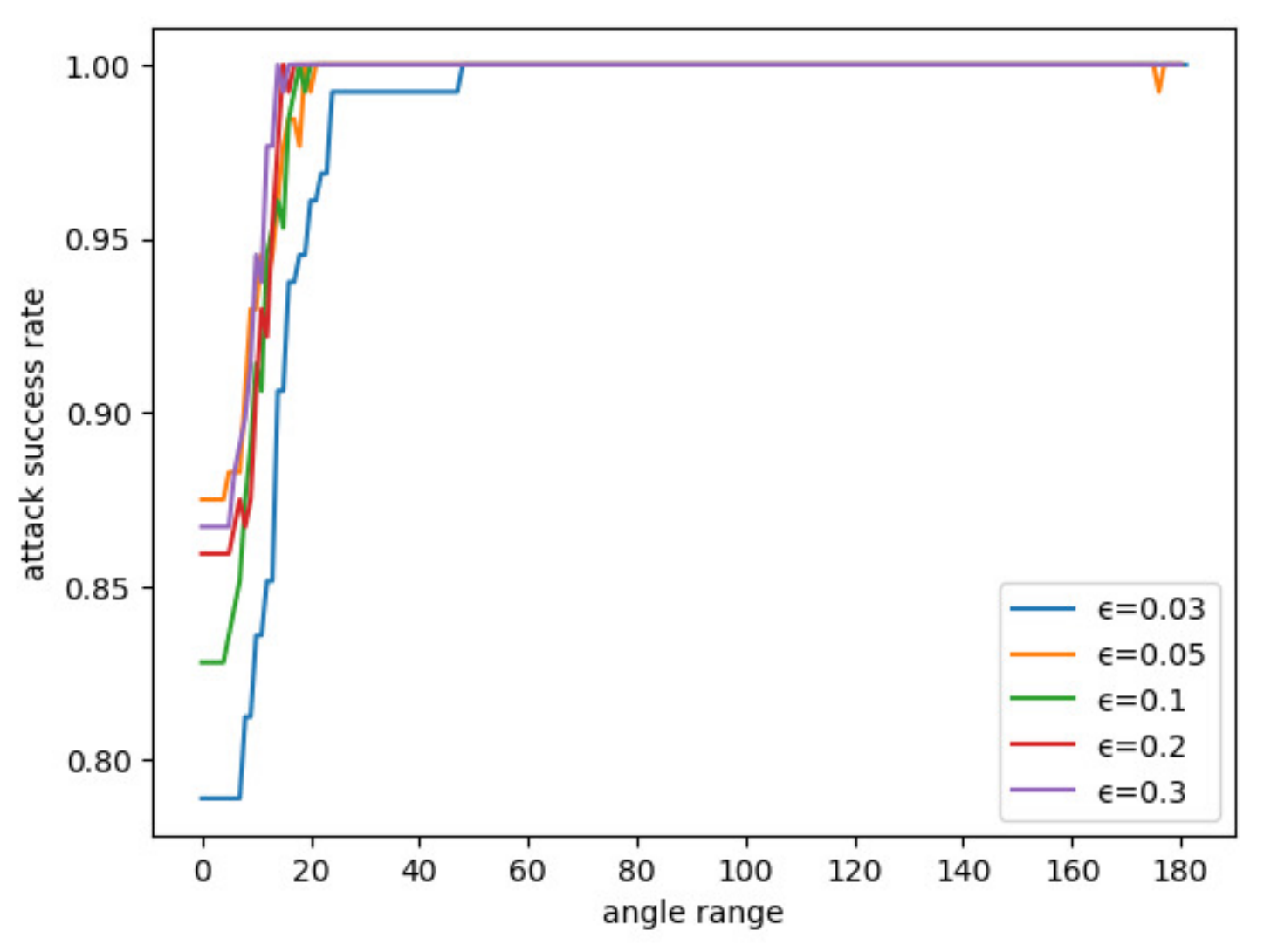}\vspace{4pt}
	\includegraphics[width=0.24\linewidth]{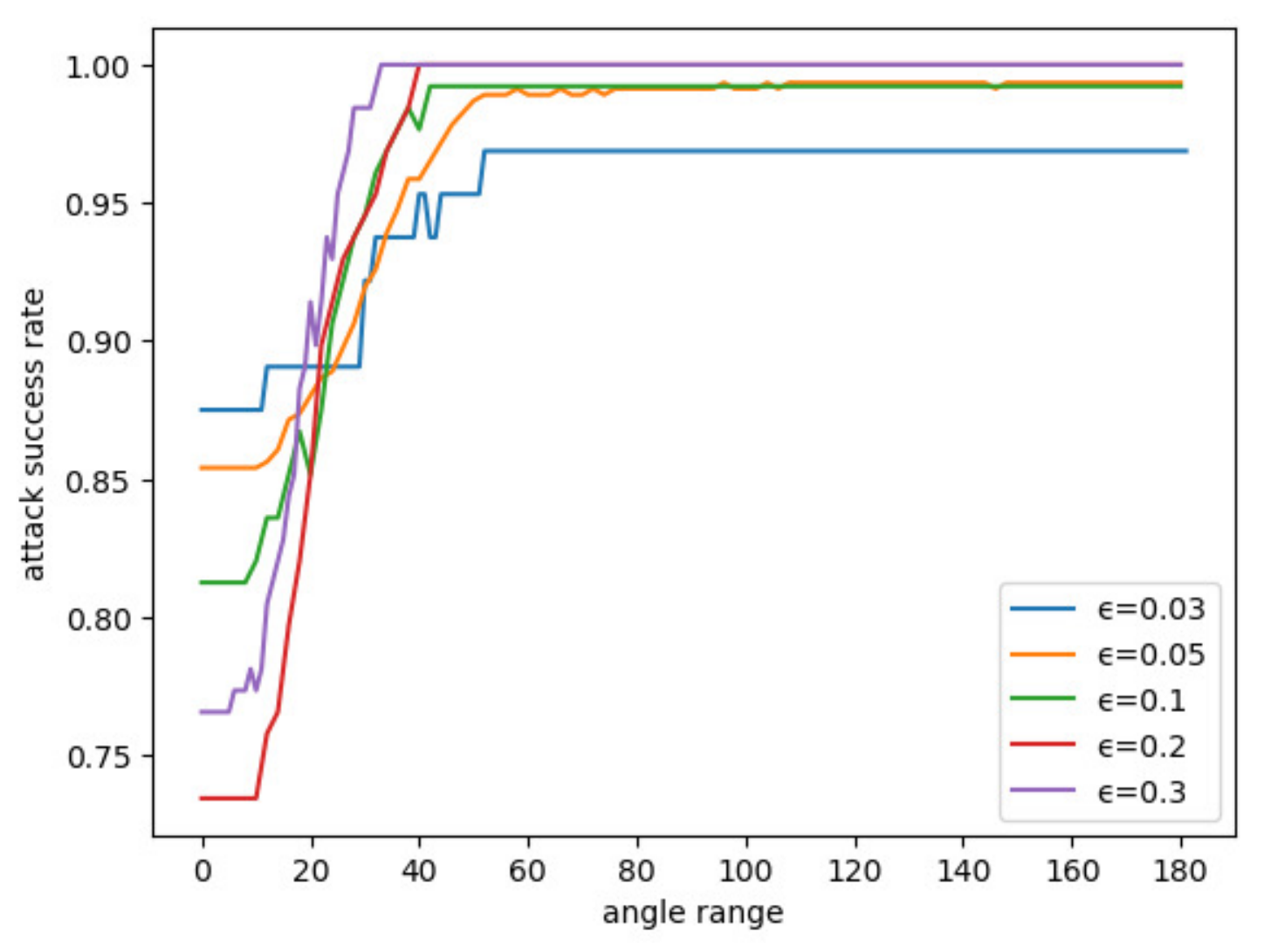}
	\caption{The impact of the selected angle on attack performance. Experimental results on MNIST, SVHN, CIFAR-10, and ImageNet-10 are given from left to right.}
	\label{angle}
\end{figure*}

The rotation angle of the RDA is $[-\theta, \theta]$, and one point is selected for every 1 degree in this range. Thus, the number of candidate directions per iteration is $2\theta$ (excluding 0 degree). 
A larger $\theta$ means that more directions can be selected, which can effectively prevent the algorithm from terminating because a local optima can be found in the failure attack direction. 
However, a larger $\theta$ also means that the number of candidate directions at each iteration increases, resulting in an increase in time complexity. 
Therefore, the size of the selected angle plays a key role in the attack performance and computational efficiency.

We test the attack performance of $\theta$ from 1 to 180 degrees, respectively, and set the number of selected dimensions $l$ in RDA to 10. The results are shown in Fig. \ref{angle}. 
Fig. \ref{angle} indicates that the attack performance first gradually increases as $\theta$ increases, and when a certain value is reached, the attack performance saturates. 
At this point, increasing $\theta$ only results in increased computational complexity but no improvement in performance. 
The performances for SVHN, CIFAR-10, and ImageNet-10 reach the upper limit when $\theta$ is about 60 degrees, and for MNIST, when $\theta$ is about 90. 

\subsubsection{Attack Direction}

\begin{figure*}
	\settoheight{\tempdima}{\includegraphics[width=.17\linewidth]{example-image-a}}%
	\centering\begin{tabular}{@{}c@{}c@{}c@{}c@{}c@{}c@{}}
		\includegraphics[width=.17\linewidth]{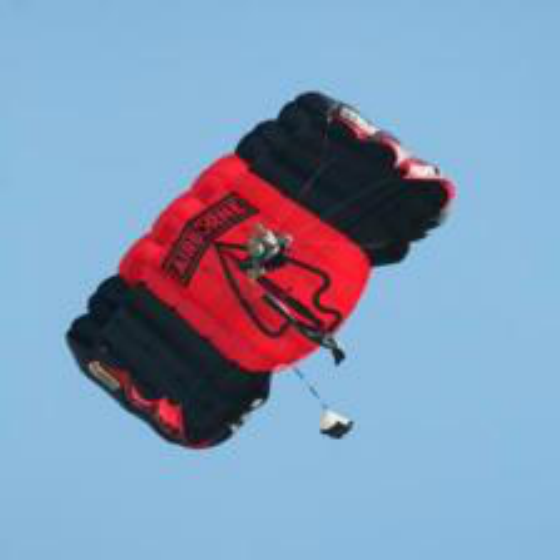}&
		\hspace{0.15cm}
		\includegraphics[width=.17\linewidth]{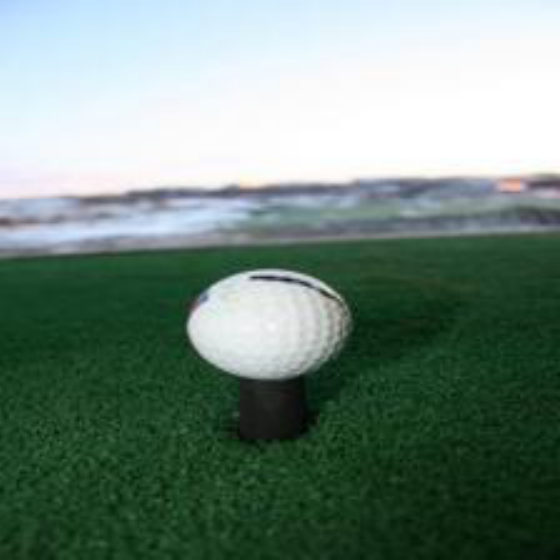}&
		\hspace{0.15cm}
		\includegraphics[width=.17\linewidth]{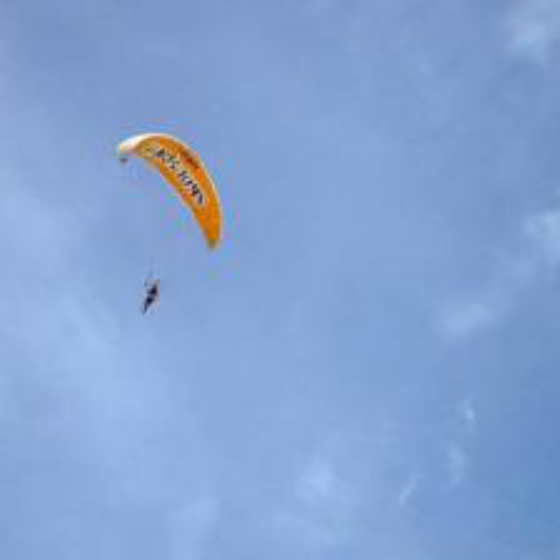}&
		\hspace{0.15cm}
		\includegraphics[width=.17\linewidth]{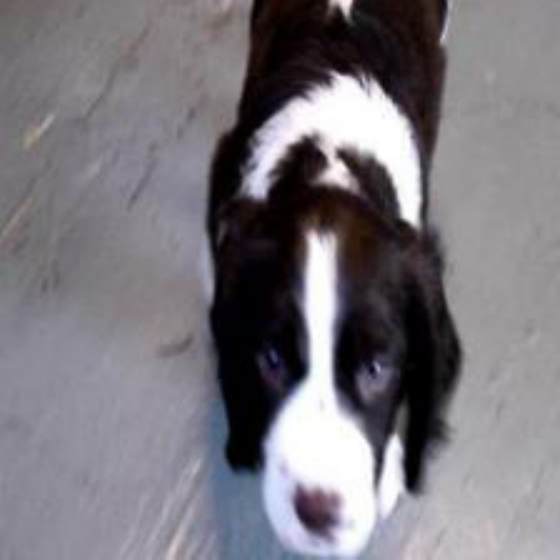}&
		\hspace{0.15cm}
		\includegraphics[width=.17\linewidth]{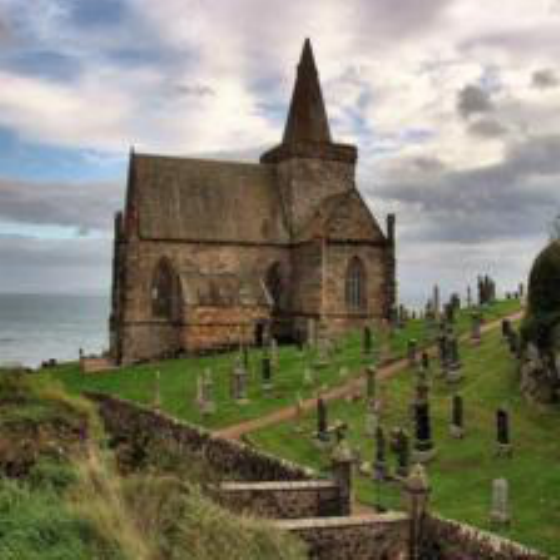}\vspace{0.1cm}\\[-1ex]
		Parachute& Golf Ball& Parachute& English Springer& Church\vspace{0.05cm}\\
		
		\includegraphics[width=.17\linewidth]{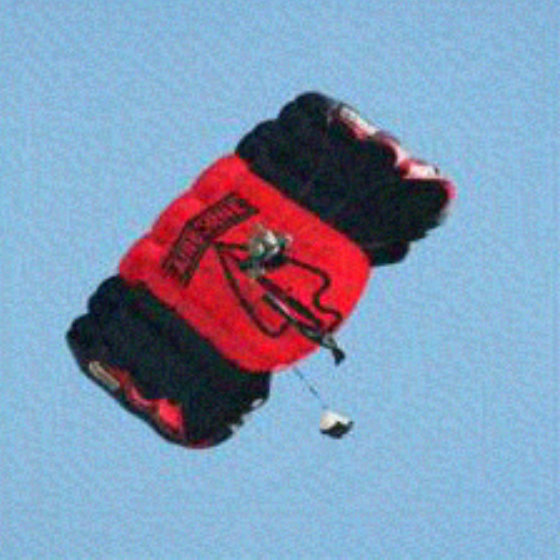}&
		\hspace{0.15cm}
		\includegraphics[width=.17\linewidth]{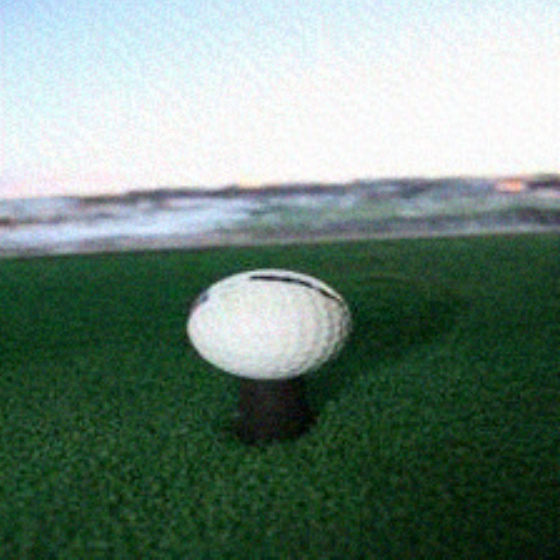}&
		\hspace{0.15cm}
		\includegraphics[width=.17\linewidth]{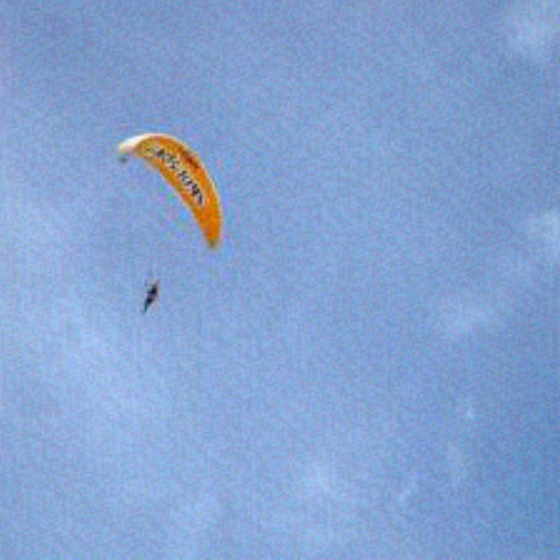}&
		\hspace{0.15cm}
		\includegraphics[width=.17\linewidth]{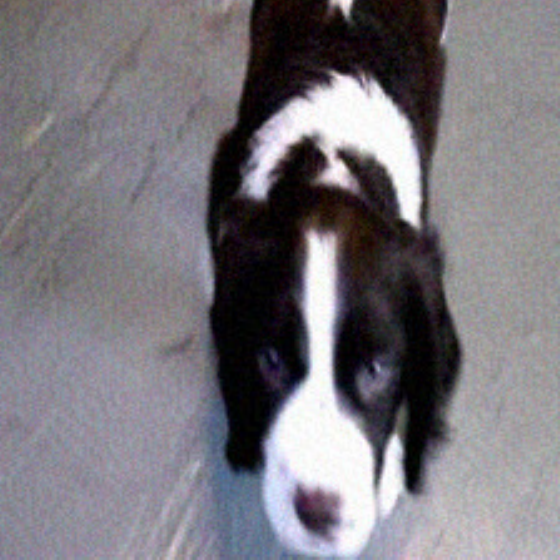}&
		\hspace{0.15cm}
		\includegraphics[width=.17\linewidth]{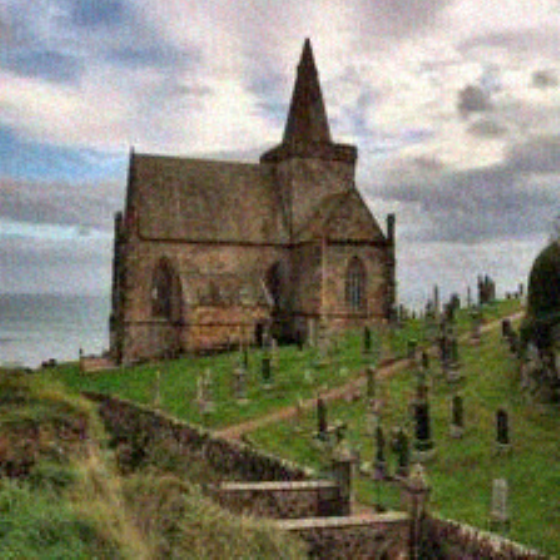}\vspace{0.1cm}\\[-1ex]
		Chain Saw ($24.7^\circ$)& Cassette Player ($32.6^\circ$)& Chain Saw ($72.0^\circ$)& Golf Ball ($88.9^\circ$)& English Springer ($90.0^\circ$)\vspace{0.1cm}\\
	\end{tabular}
	\caption{Some examples of RDA attack the classifier in ImageNet-10. The first line lists clean examples, the second line lists adversarial examples generated by the RDA ($\epsilon=0.03$). The description below each picture indicates the category predicted by the classifier and the degrees in the parentheses are the included angle between the attack direction and the gradient direction.}%
	\label{rda_example}
\end{figure*}

\begin{table*}[htbp]
	\centering
	\caption{Intersection angle of RDA direction and gradient direction}
	\begin{spacing}{1.1}
			\begin{tabular}{l|r|r|r|r|r|r|r|r|r|r|r|r}
				\hline
				\multirow{3}[0]{*}{Dataset} & \multicolumn{12}{c}{$\epsilon$} \\
				\cline{2-13}
				& \multicolumn{3}{c|}{0.03} & \multicolumn{3}{c|}{0.05} & \multicolumn{3}{c|}{0.1} & \multicolumn{3}{c}{0.2} \\
				\cline{2-13}
				& \multicolumn{1}{l|}{min} & \multicolumn{1}{l|}{max} & \multicolumn{1}{l|}{mean} & \multicolumn{1}{l|}{min} & \multicolumn{1}{l|}{max} & \multicolumn{1}{l|}{mean} & \multicolumn{1}{l|}{min} & \multicolumn{1}{l|}{max} & \multicolumn{1}{l|}{mean} & \multicolumn{1}{l|}{min} & \multicolumn{1}{l|}{max} & \multicolumn{1}{l}{mean} \\
				\hline
				MNIST & 1.79   & 90.00  & 64.68    & 0.72  & 90.00    & 70.85 & 2.86  & 90.00    & 79.77 & 1.43  & 90.00    & 85.42 \\
				\hline
				SVHN  & 2.23  & 90.00    & 52.82 & 0.70    & 90.00    & 76.40  & 0.47 & 90.00    & 31.04 & 0.47  & 90.00    & 25.50 \\
				\hline
				CIFAR-10 & 2.67  & 89.56 & 30.97 & 1.28  & 89.52 & 24.22 & 1.55  & 89.03 & 20.23 & 0.75  & 53.68 & 18.34 \\
				\hline
				ImageNet-10 &8.73       &90.00       &68.48       & 3.95  & 90.00    & 75.07 & 4.92  & 90.00    & 60.99 & 2.12  & 90.00    & 56.89 \\
				\hline
			\end{tabular}%
	\end{spacing}
	
	\label{attack_direction}%
\end{table*}%

\begin{figure*}
	\centering
	\includegraphics[width=0.24\linewidth]{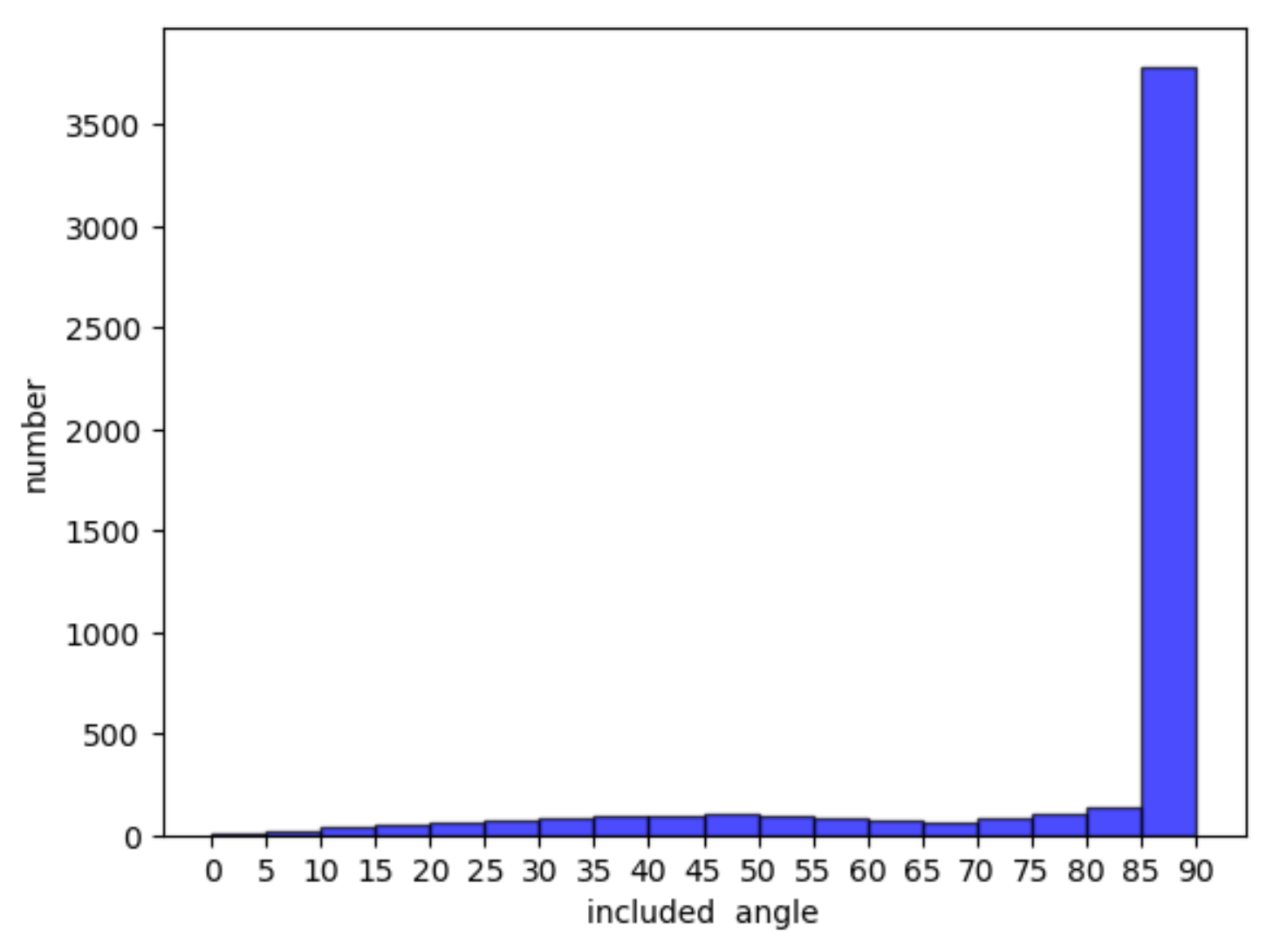}\vspace{4pt}
	\includegraphics[width=0.24\linewidth]{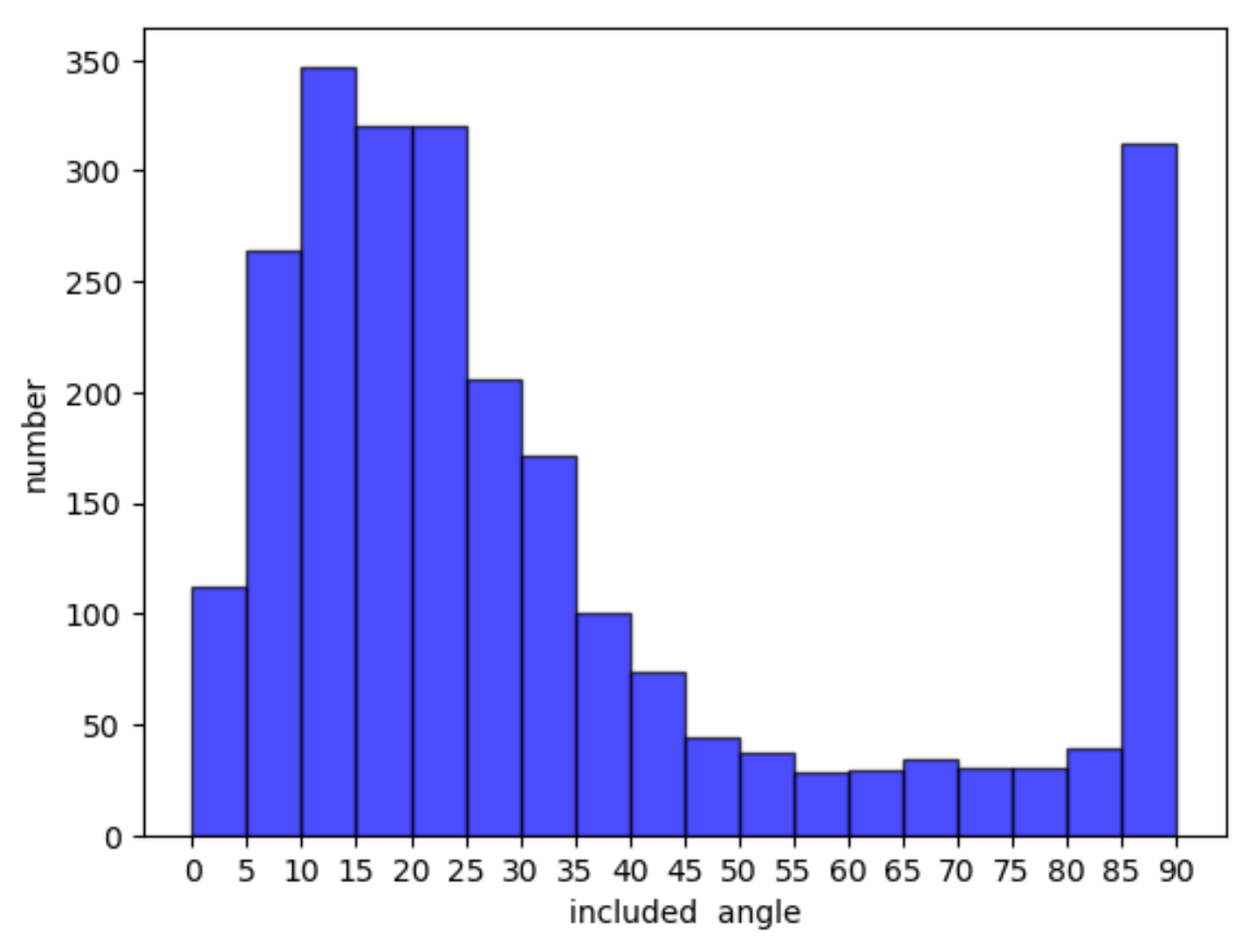}\vspace{4pt}
	\includegraphics[width=0.24\linewidth]{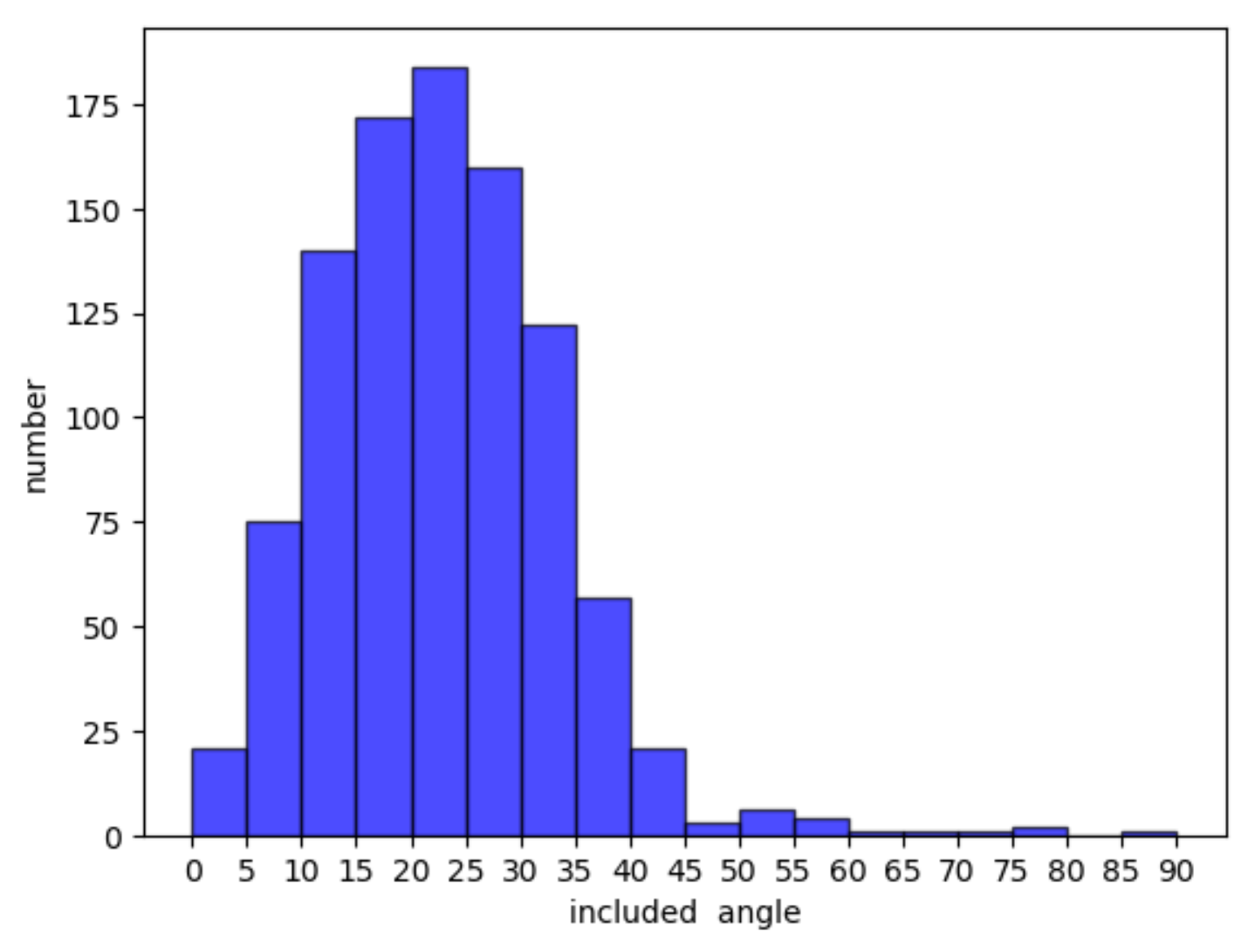}\vspace{4pt}
	\includegraphics[width=0.24\linewidth]{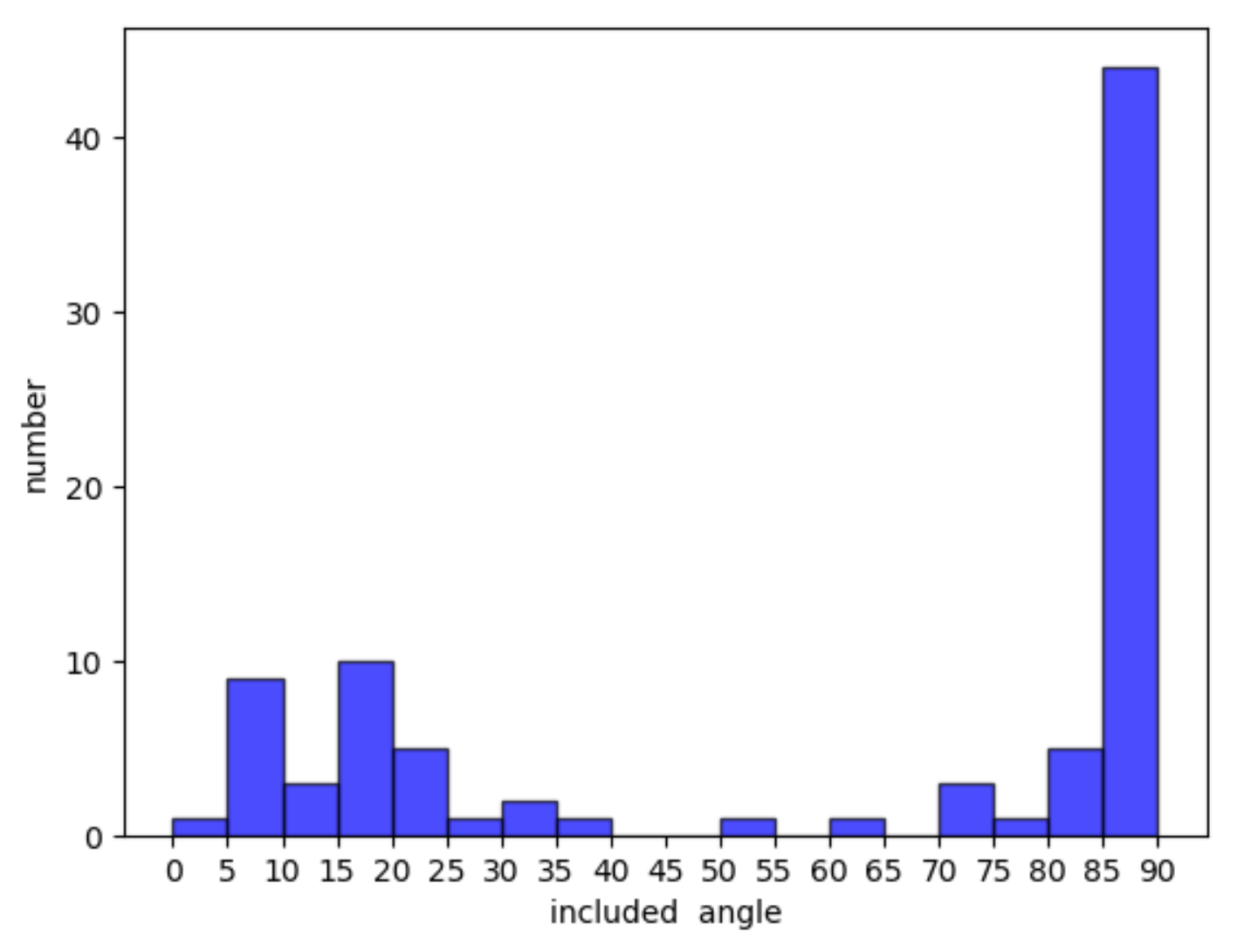}
	\caption{The distribution of the included angles  between the direction of the RDA and the gradient direction. The results from left to right are from MNIST, FMNIST, SVHN and CIFAR-10, respectively.}
	\label{angle_distribute}
\end{figure*}

The direction of RDA is often different from that of the gradient. This section experimentally calculates the included angles between the attack directions and the gradient directions to observe their relationships. 
We obtain the cosine similarity between the attack direction and the gradient direction and take the inverse cosine result as the included angle between the two directions. 

For samples that FGSM can successfully attack, the included angles between the attack direction and the gradient direction obtained by RDA mostly are very small; therefore, we only discuss the attack directions of the samples that FGSM cannot attack successfully but the RDA can. The number of selected dimensions $l$ by the RDA is 10, and the rotation angle $\theta$ in the experiments is 180 degrees. The results are summarized in Table \ref{attack_direction}.

As can be seen from Table \ref{attack_direction}, although the minimum included angle is small, the maximum included angle is $90^\circ$. The average included angle is relatively large, especially for MNIST and ImageNet-10 (above $50^\circ$). This demonstrates that effective attack directions are not always in the vicinity of the gradient direction, and there may exist a large deviation from the gradient direction. We have listed several examples of using RDA to attack ImageNet-10, as shown in Fig. \ref{rda_example}. It can also be seen from the figure that the successful attack direction of RDA does not need to be near the gradient direction.

In addition, for $\epsilon$ gradually increasing from 0.03 to 0.2, the included angle between the direction of the RDA and the gradient direction has no obvious laws. Further, we count the distribution of the included angles under the four datasets when $\epsilon$ is 0.1, as shown in Fig. \ref{angle_distribute}. We can see that the distribution of the included angles of the four datasets is also different. 

\subsubsection{Number of Iterations}
\begin{table*}[htbp]
	\centering
	\caption{Number of iterations}
	\setlength{\tabcolsep}{0.0095\linewidth}{
	\begin{tabular}{l|r|r|r|r|r|r|r|r|r|r}
		\hline
		\multicolumn{1}{c|}{\multirow{3}[0]{*}{Dataset}} & \multicolumn{10}{c}{$\epsilon$} \\
		\cline{2-11}
		& \multicolumn{2}{c|}{0.03} & \multicolumn{2}{c|}{0.05} & \multicolumn{2}{c|}{0.1} & \multicolumn{2}{c|}{0.2} & \multicolumn{2}{c}{0.3} \\
		\cline{2-11}
		& \multicolumn{1}{c|}{only\_RDAS} & \multicolumn{1}{c|}{AS} & \multicolumn{1}{c|}{only\_RDAS} & \multicolumn{1}{c|}{AS} & \multicolumn{1}{c|}{only\_RDAS} & \multicolumn{1}{c|}{AS} & \multicolumn{1}{c|}{only\_RDAS} & \multicolumn{1}{c|}{AS} & \multicolumn{1}{c|}{only\_RDAS} & \multicolumn{1}{c}{AS} \\
		\hline
		MNIST & 219.7 & 213.8 & 230.3 & 196.1 & 124.7 & 69.2 & 41.3 & 11.3 & 20.1 & 4.1 \\
		\hline
		SVHN  & 220.5 & 61.3 & 88.0 & 15.5 & 25.7 & 3.0  & 14.5 & 1.6  & 13.1 & 1.5 \\
		\hline
		CIFAR-10 & 67.0    & 10.3  & 30.8 & 3.9  & 24.8 & 3.1  & 22.6 & 2.8  & 15.1 & 1.9 \\
		\hline
		ImageNet-10 & 682.3 & 98.5 & 604.4 & 88.6  & 402.8 & 78.5 & 258.2 & 45.9 & 246.1 & 41.1 \\
		\hline
	\end{tabular}%
}
	\label{iteration}%
\end{table*}%
We already briefly analyzed the relationship between the selected dimensions and the number of iterations. 
In this section, we discuss the number of iterations required for the RDA to search for the attack direction. 
The parameter $l$ of RDA is 10, and the parameter $\theta$ is 180. We count the average iterations for all samples (AS) and the average iterations for samples (only\_RDAS) that FGSM could not successfully attack but RDA was able to successfully attack, as shown in Table \ref{iteration}.

For samples that can be successfully attacked in the gradient direction, RDA requires only a few iterations; therefore, the number of iterations of AS is smaller than only\_RDAS. 
Table \ref{iteration} indicates that, as $\epsilon$ increases, the number of required iterations decreases. This means that larger perturbations make it easier to find the feasible attack direction. 

For CIFAR-10, only\_RDAS and AS require fewer iterations. However, on MNIST and SVHN, when the $\epsilon$ is very small, the number of iterations is relatively large. On the ImageNet-10 dataset, the number of iterations in the algorithm is generally large. This is because the selected 10 dimensions are very small relative to the direction vector of 150528 dimensions. Therefore, even if the attack direction and the gradient direction are slightly different, a large number of iterations are required to arrive at a successful attack direction.

\subsection{Comparison of Results}
We compared RDA with classical gradient-based attack methods such as FGSM, L.L.Class, BIM, and MI-FGSM. We compared the attack success rate under both the white-box attack and the black-box attack. In this part of the experiments, we set $l$ to 10 and $\theta$ to 180 (it is impossible to know the specific dataset in practice; therefore, $\theta$ is set to the maximum value of 180 here).

\subsubsection{White-box Attack}
\begin{table*}[htbp]
	\centering
	\caption{Attack success rate (\%) under white-box attack}
	\setlength{\tabcolsep}{0.017\linewidth}{
		\begin{tabular}{c|l|r|r|r|r|r}
			\hline
			\multirow{2}[0]{*}{Dataset} & \multicolumn{1}{c|}{\multirow{2}[0]{*}{Attack}} & \multicolumn{5}{c}{$\epsilon$} \\
			\cline{3-7}
			&       & \multicolumn{1}{c|}{0.03} & \multicolumn{1}{c|}{0.05} & \multicolumn{1}{c|}{0.1} & \multicolumn{1}{c|}{0.2} & \multicolumn{1}{c}{0.3} \\
			\hline
			\multirow{5}[0]{*}{MNIST} & FGSM  & 2.82  & 14.86 & 44.63 & 73.42 & 81.26 \\
			& L.L.Class & 0.92  & 11.53 & 57.96 & 91.70  & 98.05 \\
			& BIM   & 7.16  & 43.87 & 92.21 & 99.95 & 99.97 \\
			& MI-FGSM & 8.16  & 58.92 & 91.97 & 99.86 & 99.98 \\
			& \textbf{RDA}   & \textbf{9.21} & \textbf{62.12} & \textbf{96.21} & \textbf{99.99} & \textbf{99.99} \\
			\hline
			\multirow{5}[0]{*}{SVHN} & FGSM  & 72.32 & 82.87 & 90.04 & 92.04 & 92.04 \\
			& L.L.Class & 75.68 & 86.99 & 99.49 & 99.90  & 99.98 \\
			& BIM   & 97.52 & 99.73 & 99.99 & \textbf{100} & \textbf{100} \\
			& MI-FGSM & 98.45 & \textbf{99.90} & \textbf{100} & \textbf{100} & \textbf{100} \\
			& \textbf{RDA}   & \textbf{98.73} & 99.89 & \textbf{100} & \textbf{100} & \textbf{100} \\
			\hline
			\multirow{5}[0]{*}{CIFAR-10} & FGSM  & 85.17 & 88.86 & 89.37 & 89.54 & 89.89 \\
			& L.L.Class & 78.48 & 83.65 & 98.21 & 99.74 & 99.82 \\
			& BIM   & 99.63 & 99.84 & \textbf{100} & \textbf{100} & \textbf{100} \\
			& MI-FGSM & 99.99 & \textbf{100} & \textbf{100} & \textbf{100} & \textbf{100} \\
			& \textbf{RDA}   & \textbf{100} & \textbf{100} & \textbf{100} & \textbf{100} & \textbf{100} \\
			\hline
			\multirow{5}[0]{*}{ImageNet-10} & FGSM  & 85.62 & 85.40  & 80.61 & 82.35 & 83.44 \\
			& L.L.Class & 93.46 & 97.39 & 96.51 & 99.13 &99.13  \\
			& BIM   & \textbf{100} & \textbf{100} & \textbf{100} & \textbf{100} & \textbf{100} \\
			& MI-FGSM & \textbf{100} & \textbf{100} & \textbf{100} & \textbf{100} & \textbf{100} \\
			& \textbf{RDA}   & 98.91 & 99.35 & 99.56 & \textbf{100} & \textbf{100} \\
			\hline
		\end{tabular}%
	}
	\label{whitebox}%
\end{table*}%
Table \ref{whitebox} lists the comparisons between the proposed RDA and the other four methods in the case of white-box attacks. 

The performance of RDA is competitive; in many cases, RDA achieves the highest attack performance. Especially, in CIFAR-10, our method successfully attacks all test samples using five sizes of perturbations. 
Although the success rate of RDA attacking MNIST is only 9.21\% when $\epsilon$ is 0.03, RDA still has the best performance compared with the other four attacks. 

Among the four methods that are compared, only BIM and MI-FGSM have attack performances similar to that of our algorithm; however, both methods are based on iterative attacks. While RDA uses multiple iterations to determine the attack direction, the attack is essentially a one-step attack.  
FGSM and L.L.Class have relatively inferior attack performances.
Although both RDA and FGSM use a one-step attack, the success rate of RDA attacks is mostly higher than FGSM by more than 10\%. 
Further, when $\epsilon$ was 0.1 on MNIST, the attack success rate increased from 44.63\% for FGSM to 96.21\% for RDA, which is the best when compared to that for all other methods. The above results indicate that RDA has excellent performance in white-box attacks.

\subsubsection{Black-box Attack}
\begin{table*}[htbp]
	\centering
	\caption{Attack success rate (\%) under black-box attack}
	\setlength{\tabcolsep}{0.017\linewidth}{
	\begin{tabular}{c|l|r|r|r|r|r}
		\hline
		\multirow{2}[0]{*}{Dataset} & \multicolumn{1}{c|}{\multirow{2}[0]{*}{Attack}} & \multicolumn{5}{c}{$\epsilon$} \\
		\cline{3-7}
		&       & 0.03  & 0.05  & 0.1   & 0.2   & 0.3 \\
		\hline
		\multirow{5}[0]{*}{MNIST} & FGSM  & 0.60  & 1.51   & 9.54 & 48.30 & 70.08 \\
		& L.L.Class & 0.70  & 6.15  & 13.52 & 36.64 & 53.47 \\
		& BIM   & 0.96  & 4.15  & 26.92 & 74.60 & 90.33 \\
		& MI-FGSM & 1.04  & 4.15  & 30.87 & 76.25  & 91.64 \\
		& \textbf{RDA}   & \textbf{8.93} & \textbf{61.05} & \textbf{95.85} & \textbf{100} & \textbf{100} \\
		\hline
		\multirow{5}[0]{*}{SVHN} & FGSM  & 46.60  & 69.35 & 87.38 & 91.57 & 91.45 \\
		& L.L.Class & 23.86 & 44.95 & 80.70  & 90.40  & 91.45 \\
		& BIM   & 52.65 & 74.03 & 95.42 & 98.28 & 98.47 \\
		& MI-FGSM & 68.65 & 87.70  & 97.02 & 98.30  & 98.34 \\
		& \textbf{RDA}   & \textbf{98.80} & \textbf{99.83} & \textbf{99.98} & \textbf{100} & \textbf{100} \\
		\hline
		\multirow{5}[0]{*}{CIFAR-10} & FGSM  & 42.15 & 63.46 & 84.96 & 89.50  & 42.15 \\
		& L.L.Class & 14.09 & 30.32 & 58.61 & 78.58 & 82.37 \\
		& BIM   & 34.15 & 60.92 & 89.73 & 96.67 & 34.15 \\
		& MI-FGSM & 51.57 & 79.35 & 94.87 & 97.40  & 51.57 \\
		& \textbf{RDA}   & \textbf{100} & \textbf{100} & \textbf{100} & \textbf{100} & \textbf{100} \\
		\hline
		\multirow{5}[0]{*}{ImageNet-10} & FGSM  & 9.37  & 22.22 & 54.90  & 75.16 & 81.92 \\
		& L.L.Class & 4.58  & 13.07 & 28.10  & 58.39 & 68.41 \\
		& BIM   & 5.01  & 14.60  & 34.20  & 61.00    & 81.05 \\
		& MI-FGSM & 9.37  & 19.61 & 49.24 & 74.95 & 68.41 \\
		& \textbf{RDA}   & \textbf{73.20} & \textbf{85.40} & \textbf{97.17} & \textbf{99.78} & \textbf{100} \\
		\hline
	\end{tabular}%
}
	\label{blackbox}%
\end{table*}%

In reality, we often cannot obtain the internal parameters of the model. The classifier often only serves as a black box and can only obtain the final classification results. 
In general, the black-box attack will train a substitute network, and then attack the target model using the adversarial samples obtained under the trained substitute model. 
RDA takes the gradient direction of the substitute model as the initial direction; however, the search for the attack direction only needs the output of the model (probability labels), and therefore, the target model can be directly used for searching in RDA. 

The final attack performance is summarized in Table \ref{blackbox}. Compared to the white-box attack, the attack performances of the four comparison methods decrease significantly, while the attack performance of RDA does not change much in most cases. 

For MNIST, SVHN, and CIFAR-10, the attack performance of RDA is comparable to that of the white-box attack. In particular, in CIFAR-10, all test samples can be successfully attacked as in the white-box attack. 

For ImageNet-10, when comparing the white-box attacks, the performance of RDA is degraded when $\epsilon$ is 0.03 and 0.05. 
However, RDA is significantly superior that the other methods as their performance decreases sharply.

Gradient-based methods such as FGSM in particular rely on the internal parameters of the model, and in some cases, the model itself is not differentiable, which limits the applications of such methods in a black-box attack. 
RDA ignores all internal features of the target model to make it effective in the more demanding black-box attack.

\section{Conclusion}
\label{sec:conclusion}
In this paper, we proposed RDA for generating adversarial samples; RDA uses the one-step attack, the attack direction is obtained using the hill climbing search, and it is not necessarily in the gradient direction. Moreover, RDA can conduct attacks with only the model output but without the internal knowledge of the target model. Therefore, under the black-box attack, RDA can achieve a similar performance as that of the white-box attack in most cases, which is difficult to achieve using other gradient-based attack methods. Although RDA is very simple, it has a very competitive attack performance when compared to the other gradient-based methods. 

The experimental results indicated that the included angles between the effective attack direction and the gradient direction vary considerably. In fact, the effective attack directions by RDA might deviate greatly from the gradient direction. In the future, to design a one-step attack, we will attempt to study more efficient methods to search the effective attack direction.

\bibliographystyle{IEEEtran}
\bibliography{ref}

\vfill

\end{document}